\documentclass[%
reprint,
superscriptaddress,
amsmath,
amssymb,
prd,
]{revtex4-2}

\usepackage{graphicx}
\usepackage{subfigure}
\usepackage{url,hyperref} 
\usepackage{color}
\usepackage{bm}

\usepackage{natbib}

\newcommand{\wt}[1]{\widetilde{#1}}

\begin{document}
\title{Charged particle beam transport in a flying focus pulse with orbital angular momentum}
\author{Martin Formanek}
\email{martin.formanek@eli-beams.eu}
\affiliation{Max Planck Institute for Nuclear Physics, Saupfercheckweg 1, D-69117 Heidelberg, Germany}
\affiliation{ELI Beamlines Facility, The Extreme Light Infrastructure ERIC, 252 41 Doln\'{i} B\v{r}e\v{z}any, Czech Republic}
\author{John P. Palastro}
\affiliation{Laboratory for Laser Energetics, University of Rochester, Rochester, New York 14623, USA}
\author{Marija Vranic}
\affiliation{GOLP/Instituto de Plasma e Fus\~{a}o Nuclear, Instituto Superior T\'ecnico, Universidade de Lisboa, 1049-001 Lisbon, Portugal}
\author{Dillon Ramsey}
\affiliation{Laboratory for Laser Energetics, University of Rochester, Rochester, New York 14623, USA}
\author{Antonino Di Piazza}
\affiliation{Max Planck Institute for Nuclear Physics, Saupfercheckweg 1, D-69117 Heidelberg, Germany}

\date{\today}
%
\begin{abstract}
We demonstrate the capability of Flying Focus (FF) laser pulses with $\ell = 1$ orbital angular momentum (OAM) to transversely confine ultra-relativistic charged particle bunches over macroscopic distances while maintaining a tight bunch radius. A FF pulse with $\ell = 1$ OAM creates a radial ponderomotive barrier that constrains the transverse motion of particles and travels with the bunch over extended distances. As compared to freely propagating bunches, which quickly diverge due to their initial momentum spread, the particles co-traveling with the ponderomotive barrier slowly oscillate around the laser pulse axis within the spot size of the pulse. This can be achieved at FF pulse energies that are orders of magnitude lower than required by Gaussian or Bessel pulses with OAM. The ponderomotive trapping is further enhanced by radiative cooling of the bunch resulting from rapid oscillations of the charged particles in the laser field. This cooling decreases the mean square radius and emittance of the bunch during propagation.
\end{abstract}
\maketitle


\section{Introduction}
Charged particle beams are ubiquitous in physics experiments and applications. Their transport over macroscopic distances is necessary not only for devices like accelerators and electron microscopes, but also for compact radiation sources. The magnetic optics currently used for transport \cite{weingartner2011imaging,andre2018control} become progressively more expensive as the particle energy increases due to the need for higher magnetic field gradients, which must be generated, for example, by superconducting currents. Further, the achievable focal lengths of these optics may be too large for modern electron beam applications, such as inverse Compton scattering sources \cite{lim2005adjustable}. More specifically, the characteristic focusing lengths of magnetic optics range from tens of centimeters to meters and are limited by the physical size of the magnets. While permanent magnets designed for $\sim$100 MeV electron energies can be relatively compact (millimeter scale) allowing for $\sim$10 cm focal lengths \cite{lim2005adjustable}, at higher electron energies, focusing at such short distances becomes technologically challenging.

\begin{figure}
	\includegraphics[width=\linewidth]{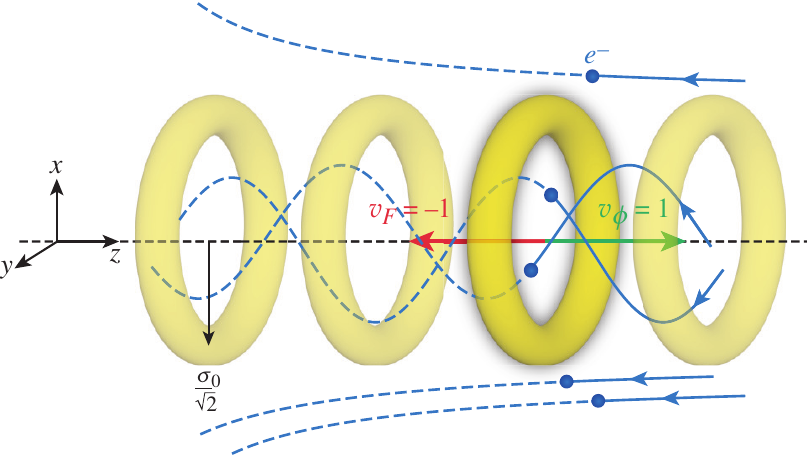}
	\caption{\label{fig:graphics} Schematic of electron confinement in the ponderomotive potential of an $\ell = 1$ OAM flying focus (FF) pulse. The off-axis intensity peak of the FF pulse (yellow toroids) travels at the vacuum speed of light ($v_F = -1$) in the opposite direction of its phase fronts ($v_\phi = 1$). Ultrarelativistic electrons (blue lines) travel in the same direction as the intensity peak. Electrons with low transverse momentum inside the intensity peak are confined and slowly oscillate in the radial direction, whereas electrons outside the peak are deflected. [The solid (dashed) blue lines represent the past (future) trajectory of the electrons with respect to the snapshot shown in the figure].}
\end{figure}	

All-optical setups for transporting charged particle beams have been proposed as an alternative that would circumvent the need for magnets \cite{Bialynicki-Birula:2004bvr}. In these schemes, the transverse intensity profile of a laser pulse that counterpropagates with respect to the beam is shaped to provide a confining ponderomotive potential. Such schemes can be especially advantageous when employed at high-intensity laser facilities, where laser pulses can be used both for creating ultra-relativistic particle beams and for their transport. Nevertheless, these schemes require large laser pulse energies, whether they employ conventional or axicon-focused Bessel pulses \cite{mellado2006trapping,bialynicki2012trapping,Schachter:2020jaw}.

In this work, we introduce an all-optical setup for charged particle beam guiding that uses a flying focus (FF) to greatly reduce the required laser pulse energy. The FF refers to a class of optical techniques that provide spatiotemporal control over the trajectory of a focal point \cite{Sainte-Marie_2017,Froula_2018}. The intensity peak formed by the moving focal point can travel at any arbitrary velocity independent of the laser group velocity over distances much longer than a Rayleigh range. The first experimental demonstration of a FF used chromatic focusing of a chirped laser pulse \cite{Froula_2018}. Alternate techniques employing space-time light sheets \cite{kondakci2017diffraction,yessenov2020accelerating}, axiparabola-echelon mirrors \cite{Palastro_2020}, and nonlinear media \cite{simpson2020nonlinear,simpson2022spatiotemporal} have also been proposed. The spatiotemporal control enabled by FF pulses has presented a unique opportunity to revisit established schemes and investigate regimes in which FF pulses provide an advantage over traditional fixed-focus Gaussian pulses \cite{Turnbull_2018,Palastro_2018,Howard_2019,Palastro_2020,Ramsey_2020,DiPiazza:2020wxp,Ramsey_2022,Formanek_2022}. 

Here we show that FF pulses with $\ell = 1$ orbital angular momentum (OAM) and a focus that moves in the opposite direction of the phase fronts at the vacuum speed of light can transport charged particles over macroscopic distances. The ring-shaped transverse intensity profile of the $\ell = 1$ mode provides a ponderomotive potential barrier that confines charged particles in the transverse direction. Figure \ref{fig:graphics} illustrates the concept and problem geometry (the units $\hbar = \varepsilon_0 = c = 1$ are used throughout). The ponderomotive confinement allows for transport of an electron bunch with a tight radius, smaller that the focal spot size of the pulse, with feasible laser pulse energies. As an example, a 10 pC, 500 MeV electron beam can be transported over 6 mm using a 200 J, 5 TW FF pulse, compared to the 2 MJ that would be needed in a conventional pulse. Axicon-focused Bessel pulses require even larger energies \cite{Schachter:2020jaw,schachter2022normalized}. The energy requirements to confine the electron beam are much less for a FF pulse because the peak intensity moves with the electron beam, which decouples the interaction length from the Rayleigh range. Confinement of the charged particles is aided by radiative cooling (radiation reaction \cite{di2012extremely,gonoskov2022charged,fedotov2022advances,di2022multi}), which decreases the emittance and mean-squared radius of the beam at the cost of its average energy.

The remainder of the article is organized as follows: Section \ref{sec:fffields} presents the four-potential of the $\ell = 1$ OAM FF beam and its properties. Section \ref{sec:motion} describes the analytical model for charged particle motion in the $\ell = 1$ FF pulse. In Section \ref{sec:ponderomotive}, oscillations in the ponderomotive potential are discussed and constraints on the transverse phase space of the trapped particles are derived. Section \ref{sec:RMS} describes the evolution of the particle bunch radius. Longitudinal motion of the particles, including radiation energy loss, is addressed in Section \ref{sec:longitudinal}. The importance of space charge repulsion on particle bunches is discussed in Section \ref{sec:coulomb_text}. Section \ref{sec:energy_estimates} compares the energy required in $\ell = 1$ FF pulses and conventional Laguerre-Gaussian pulses. Section \ref{sec:simulations} describes the simulation results. Electron confinement, radiation energy loss, oscillations in the ponderomotive potential, and transverse emittance behavior are covered in Section \ref{sec:radiation_cooling}, and the longitudinal delay of the electron bunch behind the FF intensity peak is covered in Section \ref{sec:longitudinal_simulations}. Technical details and explicit calculations are contained in six appendices.
\section{FF beams and pulses with $\ell = 1$ OAM}\label{sec:fffields}
A FF field with $\ell = 1$ OAM forms a transverse ponderomotive potential barrier that is capable of confining charged particles close to the propagation axis. Here, a FF pulse with an intensity peak that moves at the vacuum speed of light against the laser phase velocity in the negative $z$-direction (see Fig. \ref{fig:graphics}) is considered. This special case admits simple, but exact, closed-form analytical expressions for the four-potential and the fields. This section presents the exact expressions of the four-potential, its cycle-averaged magnitude, and the extension to pulses with finite energy. A scheme for generating FF pulses with $\ell = 1$ OAM is described in Ref. \cite{simpson2022spatiotemporal}.

An exact beam solution to the vacuum wave equation can be written in terms of the lightcone coordinates $\eta = t + z$ and $\phi = t - z$. The lightcone coordinate $\eta$ describes the displacement from the moving focus ($\eta = 0$) of the FF pulse, while $\phi$ tracks the fast phase oscillations. The transverse part of the vector potential for the $\ell = 1$ Laguerre-Gaussian (LG10) mode reads \cite{Formanek_2022}
\begin{equation}\label{eq:aperp}
	\bm{A}_{\perp}(\eta,r,\theta,\phi) = \bm{\mathcal{A}}_0 \frac{\sqrt{2}\sigma_0 r}{\sigma^2_\eta} e^{-r^2/\sigma^2_\eta} \cos\Psi_1(0,0)\,,
\end{equation}
where
\begin{equation}\label{eq:phase}
	\begin{split}
		\Psi_1(a,b) = \omega_0 \phi -  &\frac{r^2}{\sigma^2_\eta}\frac{\eta}{\eta_0} + (1-a) \theta\\
		&+ (2 + b)\arctan\left(\frac{\eta}{\eta_0}\right)
	\end{split}
\end{equation}
is the phase, $r=|\bm{x}_\perp|=\sqrt{x^2+y^2}$ is the radial distance from the $z$-axis, $\theta = \arctan(y/x)$ is the azimuthal angle in cylindrical coordinates, $\omega_0 = 2\pi/\lambda_0$ is the laser angular frequency, and $\lambda_0$ its wavelength. The spot size $\sigma_\eta$ and Rayleigh range $\eta_0$ equivalents for the FF beam are
\begin{equation}\label{eq:rayleigh_range}
	\sigma_\eta = \sigma_0\sqrt{1+\frac{\eta^2}{\eta_0^2}}, \quad \eta_0 = \omega_0 \sigma_0^2\,.
\end{equation}
The effective duration of the moving intensity peak is equal to the Rayleigh range $\eta_0$. More generally, the Rayleigh range and effective duration depend on the velocity of the focus $\beta_F$. In the paraxial approximation, $\eta_0 = (1-\beta_F)\omega_0 \sigma_0^2 /2$, which reduces to Eq. \ref{eq:rayleigh_range} when $\beta_F = -1$ and the stationary focus result $\eta_0 =\omega_0 \sigma_0^2 /2 $ when $\beta_F = 0$ \cite{simpson2022spatiotemporal,Ramsey_2022b}. 

Upon imposing the Lorenz gauge condition $\partial_\mu A^\mu = 0$ and the constraint $A_{+} = A^0 + A^z = 0$ (so that two photon degrees of freedom remain), one can evaluate $A_{-} = A^0 - A^z$ as
\begin{equation}
	A_{-}(\eta,r,\theta,\phi) = - \int d\phi \bm{\nabla}_\perp \cdot \bm{A}_{\perp}(\eta,r,\theta,\phi)\,.
\end{equation}
For a laser beam polarized along the $y$-axis $\bm{\mathcal{A}}_0 = \mathcal{A}_0 \hat{\bm{y}}$
\begin{equation}
	\begin{split}\label{eq:aminus}
		A_{-}(\eta,r,&\theta,\phi) = \mathcal{A}_0 \frac{\sqrt{2}\sigma_0 r}{\omega_0\sigma^2_\eta} e^{-r^2/\sigma^2_\eta}\\ &\times\left[\frac{2y}{\sigma_0 \sigma_\eta} \sin \Psi_1(0,1) - \frac{1}{r} \cos \Psi_1(1,0) \right]\,,
	\end{split}
\end{equation}
where the initial condition at $t = 0$ was chosen so that the potential vanishes as $|z|\rightarrow \infty$. The remaining Cartesian components can be evaluated as
\begin{equation}\label{eq:A0}
	A^0 = - A^z = \frac{1}{2}A_{-}\,.
\end{equation}
In this gauge, the Lorentz-invariant square of the four-potential is given by
\begin{equation}
	A_{\mu} A^\mu = (A^0)^2 - (A^y)^2 - (A^z)^2 = -A_\perp^2\,, 
\end{equation}
where the Minkowski metric tensor is chosen as $\eta_{\mu\nu} = \text{diag}(+1,-1,-1,-1)$. As a result, the square of the transverse component $|A_\perp|^2 = - A_\mu A^\mu$ is also Lorentz invariant. Components of the electric and magnetic fields corresponding to this four-potential are presented in Appendix \ref{sec:emfields}. 

The cycle-average of the invariant $|A_\mu A^\mu|$ at focus ($\eta = 0$) is proportional to intensity and equal to
\begin{equation}\label{eq:aperpave}
	\left.\overline{|A_\mu A^\mu|}\right|_{\eta = 0} = \overline{A_\perp^2}|_{\eta = 0} = \frac{\mathcal{A}_0^2 r^2}{\sigma_0^2}e^{-2r^2/\sigma_0^2}\,,
\end{equation}
where the overbar denotes a cycle-average. The cycle averaging procedure is defined in Appendix \ref{sec:beam_power}. Figure \ref{fig:A2ave} displays Eq. (\ref{eq:aperpave}) and demonstrates that the peak intensity forms a ring surrounding the beam axis.
\begin{figure}
	\includegraphics[width=\linewidth]{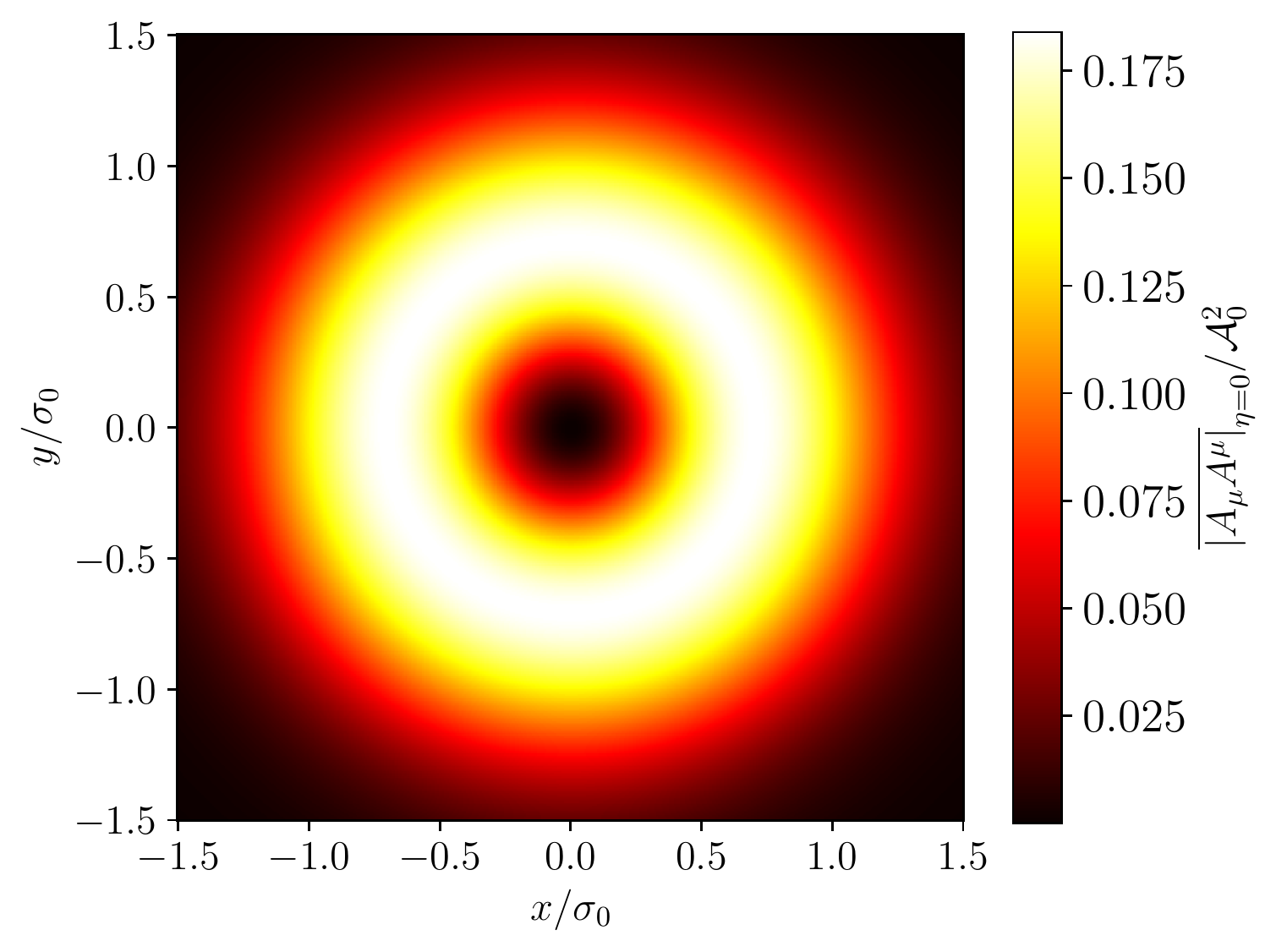}
	\caption{\label{fig:A2ave} Cycle-averaged invariant $|A_\mu A^\mu|$ at the position of the focus ($\eta = 0$).}	
\end{figure}

From the analytical beam solution, laser pulses with finite total energy can be approximately constructed by applying the pulse envelope function $g(\phi)$ as a multiplicative factor on the electromagnetic fields. For the approximation to be accurate, the up and down ramps of $g(\phi)$ should be much longer than $\lambda_0$. For details of the implementation, see Refs. \cite{DiPiazza:2020wxp,Formanek_2022}. 

The average power $P_{\text{ave}}$ of the FF pulse is given by (see Appendix \ref{sec:beam_power})
\begin{equation}\label{eq:avepower}
	P_\text{ave} \approx \frac{\pi}{4}\mathcal{A}_0^2 \omega_0^2 \sigma_0^2\,.
\end{equation}
To ensure that the FF pulse interacts with the particles for a time $t_\text{int}$, or equivalently a length $L_\text{int}$, the total pulse energy must be
\begin{equation}\label{eq:totenergy}
	E_\text{tot} \approx 2 P_\text{ave} t_\text{int}\,.
\end{equation}
Comparison with the energy required in a conventional LG10 pulse is presented in Section \ref{sec:energy_estimates}. 
\section{Charged particle motion in an $\ell = 1$ OAM FF pulse}\label{sec:motion}
The evolution of a charged particle (mass $m$ and charge $q$, respectively) interacting with an external electromagnetic field, including radiation reaction in the classical regime, is described by the Landau-Lifshitz equation of motion  \cite{Landau_b_2_1975}: 
\begin{equation}\label{eq:LL}
	\begin{split}
		\dot{u}^\mu &= \frac{q}{m}F^{\mu\nu}u_\nu \\ 
		&+\frac{2q}{3m}r_q\left[\frac{d}{d\tau}(F^{\mu\nu})u_\nu + \frac{q}{m}P^\mu_\nu F^{\nu\alpha}F_{\alpha\beta}u^\beta\right],
	\end{split}
\end{equation}
where $r_q = q^2/(4\pi m)$ is the classical particle radius, $\dot{u}^\mu$ denotes the proper-time derivative of the four-velocity $u^\mu = (\gamma,\bm{u}) = (\gamma,\gamma \bm{\beta})$, $P^\mu_\nu = \delta^\mu_\nu - u^\mu u_\nu$ is the projection tensor, and $F^{\mu\nu} = \partial^\mu A^\nu - \partial^\nu A^\mu$ is the electromagnetic field tensor in terms of the vector four-potential.

For the analytic considerations in this section, radiation reaction is assumed to be negligible, allowing the term proportional to $r_q$ to be omitted. This term has an effect of lowering the particle energy during the propagation and will be revisited in the context of the longitudinal motion (Section \ref{sec:longitudinal}) and simulation results (Section \ref{sec:simulations}). 

In the absence of radiation reaction, the motion of a charged particle in the FF pulse can be described by the ponderomotive guiding center equation of motion \cite{Quesnel:1998zz}
\begin{equation}\label{eq:quesnel}
	\frac{d\overline{\bm{u}}}{dt} = - \frac{q^2}{2m^2\overline{\gamma}}\bm{\nabla} \overline{A_\perp^2}\,,
\end{equation}
where $\overline{\gamma} = (1 + \overline{u}_z^2 + \overline{\bm{u}}_\perp^2 + q^2 \overline{A^2_\perp}/m^2)^{1/2}$ is the cycle-averaged gamma factor and $t$ is laboratory time. Equation \ref{eq:quesnel} was derived in the Coulomb gauge and requires that the particle experiences many phases of the field over the duration of the pulse, i.e., $\eta_0\omega_0(1-\beta_z) \gg 2\pi$, which is the case considered here. While there is no consensus on a covariant and gauge-independent formulation of the ponderomotive force \cite{manheimer1985covariant, startsev1997multiple}, Eq. \ref{eq:quesnel} accurately predicts the motion independent of gauge as long as the cycle averaging procedure remains valid. In Appendix \ref{sec:ponderomotive_app}, the transverse component of the ponderomotive force is derived in the Lorenz gauge for ultrarelativistic particles moving with the intensity peak ($\eta \approx 0$) and against the phase fronts of a FF pulse. The particles have an initial relativistic factor $\gamma_0 \gg 1$ and a small transverse velocity $\xi_0 \ll \gamma_0$, such that $\overline{\gamma}\approx \gamma_0$, which further simplifies the expression for $\overline{\gamma}$ in Eq. (\ref{eq:quesnel}). Applying these conditions to Eq. (\ref{eq:quesnel}) yields the same result as Eq. (\ref{eq:ponderomotive_app}):
\begin{equation}\label{eq:ponderomotive}
	\frac{d\overline{\bm{u}}_\perp}{dt} \approx - \frac{q^2}{2m^2\gamma_0} \bm{\nabla}_\perp \overline{A_\perp^2}|_{\eta = 0}\,.
\end{equation}
For the remainder of the manuscript, the velocity components and spatial coordinates will be understood to represent cycle averaged quantities, and the overbar will be dropped.

Near the focus ($\eta\approx 0$), $\overline{A_\perp^2}$ is only a function of the radial coordinate [see Eq. (\ref{eq:aperpave})]. As a result, the motion in the transverse plane is approximately described by
\begin{align}
	\label{eq:ddr}r'' - r (\theta')^2&= - \frac{q^2}{2m^2\gamma_0^2} \frac{d}{dr} \overline{A_\perp^2}|_{\eta = 0}\,,\\
	\left(\gamma_0 r^2 \theta' \right)' &= 0\,,	
\end{align}
where prime denotes a derivative with respect to time in the laboratory frame. The first equation describes the radial motion, and the second equation implies the conservation of relativistic angular momentum
\begin{equation}
	L_z = m \gamma_0 r^2 \theta' = \text{const.}
\end{equation}
Using this constant of motion in Eq. (\ref{eq:ddr}) provides
\begin{equation}\label{eq:radial}
	r'' - \frac{L_z^2}{m^2 \gamma_0^2 r^3} = -\frac{q^2}{2m^2\gamma_0^2} \frac{d}{dr} \overline{A_\perp^2}|_{\eta = 0}\,.
\end{equation}
Equation (\ref{eq:radial}) can be re-expressed in the form of Newton's law for a particle with a mass $m\gamma_0$ moving in a potential that depends only on the radial coordinate:
\begin{equation}\label{eq:newton}
	m \gamma_0 r'' = - \frac{d}{dr}V_\text{eff}(r)\,,
\end{equation}
where the effective potential
\begin{equation}\label{eq:potential}
	\begin{split}
	V_\text{eff}(r) &= V_P(r) + V_C(r)\\
	&=\frac{q^2}{2m\gamma_0}\overline{A_\perp^2}|_{\eta = 0} + \frac{L_z^2}{2m\gamma_0 r^2}
	\end{split}
\end{equation}
includes both the ponderomotive and centrifugal contributions. 
\subsection{Transverse motion}\label{sec:ponderomotive}
To illustrate the approximate harmonic motion of charged particles in the FF pulse, Eq. (\ref{eq:ponderomotive}) can be rewritten using the potential from Eq. (\ref{eq:aperpave}) as
\begin{equation}
\label{eq:accel_ponderomotive}
\frac{d^2\bm{x}_{\perp}}{dt^2}=-\Omega^2\left(1-2\frac{r^2}{\sigma_0^2}\right)e^{-2r^2/\sigma_0^2}\bm{x}_{\perp},
\end{equation}
where $\xi_0 = |q|\mathcal{A}_0 / m$ is the normalized field amplitude and
\begin{equation}\label{eq:period}
\Omega = \frac{2\pi}{T} = \frac{\xi_0}{\gamma_0 \sigma_0}
\end{equation} 
is the angular frequency of oscillations in the harmonic approximation, i.e., when $r \ll \sigma_0$. The actual oscillation frequency of confined particles is smaller due to the anharmonicity of the potential.

Multiplying Eq. (\ref{eq:newton}) by the radial velocity $r' = \beta_r$ and integrating over time provides a conservation relation for the energy $\mathcal{E}_\perp$ associated with the transverse motion:
\begin{equation}\label{eq:energyconserv}
	\frac{1}{2}m\gamma_0\beta_r^2 + V_\text{eff}(r) = \frac{1}{2}m\gamma_0 \beta_\perp^2 + V_P(r) = \mathcal{E}_\perp\,,
\end{equation}
where
\begin{equation}
	\beta_\perp = \sqrt{(r')^2 + r^2 (\theta')^2} = \sqrt{\beta_x^2 + \beta_y^2}
\end{equation}
is the magnitude of the transverse velocity. In terms of the Cartesian velocities and positions, $r' = (x\beta_x + y \beta_y)/r$.

Equation (\ref{eq:energyconserv}) can be used to determine the initial conditions of particles that will be bound in the FF potential. To begin, note that $V_\text{eff}(r) \rightarrow \infty$ as $r \rightarrow 0$ and $V_\text{eff}(r) \rightarrow 0$ as $r \rightarrow \infty$. Because there cannot be bound trajectories in regions of space where the potential is monotonically decreasing, $dV_\text{eff}(r)/dr\ge 0$ provides a necessary condition for the existence of bound trajectories. Upon applying this inequality to Eq. \ref{eq:potential}, one obtains
\begin{equation}
\label{eq:max_min}
\rho^2\le \frac{r^4}{\sigma_0^4}\left(1-2\frac{r^2}{\sigma_0^2}\right)e^{-2r^2/\sigma_0^2}\lessapprox 0.02,
\end{equation}
where $\rho=L_z/(m\sigma_0\xi_0)$ and 0.02 is a numerical upper bound on the RHS. Physically, a particle with too large of an angular momentum will not be bound in the ponderomotive potential. 

Equality in Eq. (\ref{eq:max_min}) determines the local extrema of $V_\text{eff}(r)$. For $\rho \ll 1$, the position $r_\text{max}$ of the local maximum of $V_\text{eff}(r)$ can be approximated to leading order as
\begin{equation}
r_{\text{max}} = \frac{\sigma_0}{\sqrt{2}}\,.
\end{equation}
The position $r_\text{min}$ of the local minimum of $V_\text{eff}(r)$ is obtained by assuming $r_\text{min} \ll \sigma_0$. To leading order
\begin{equation}
	r_\text{min} = \sigma_0  \sqrt{\rho}\,.
\end{equation}
Consistent with the expression for the effective potential, a non-zero initial angular momentum prevents the particle from penetrating the potential all the way to $r = 0$.

Using these values of the extrema, bound trajectories in the FF beam are determined by the constraints $V_{\text{eff}}(r_{\text{min}})\le \mathcal{E}_{\perp}\le V_{\text{eff}}(r_{\text{max}})$, i.e.,
\begin{equation}
\label{eq:phasespace}
4e\rho\le \left(\frac{\beta_\perp}{\beta_{\perp,\max}}\right)^2 + \left(\sqrt{e}\frac{r}{r_\text{max}} e^{-r^2/\sigma_0^2}\right)^2 \leq 1\,,
\end{equation}
where $\beta_{\perp,\max}=\xi_0/(\sqrt{2e}\gamma_0)$ and $e = 2.7183\ldots$ is Euler's number. Note that for this derivation to be consistent $\sqrt{\rho}\ll 1$ and $4e\rho< 1$.

\subsection{Evolution of the particle bunch in the harmonic approximation}\label{sec:RMS}
The previous section described the radial dynamics of individual particles. In this section, the dynamics of a particle bunch is described in terms of the bunch centroid in the transverse plane $\langle \bm{x}_\perp \rangle$ and the root mean squared (RMS) radius of the bunch $R = \sqrt{\langle r^2 \rangle}$. Here the ensemble average of a quantity $Q$ is defined as $\langle Q\rangle = N^{-1}\sum_{i=1}^N Q_i$, where $N$ is the number of particles. 

In the harmonic approximation, the centroid evolves according to harmonic part of Eq. (\ref{eq:accel_ponderomotive}) averaged over an ensemble of particles
\begin{equation}\label{eq:centroid}
	\frac{d^2\langle \bm{x}_\perp \rangle}{dt^2} = -\Omega^2 \langle \bm{x}_\perp \rangle\,.
\end{equation}
Equation \ref{eq:centroid} has the solution $\langle \bm{x}_\perp(t) \rangle = \langle \bm{x}_\perp(0) \rangle \cos(\Omega t)$. Thus an electron bunch that is initially offset from the propagation axis of the FF pulse will oscillate about the axis with a period $2\pi/\Omega.$

An evolution equation for $R$ can be derived by taking its second derivative with respect to the laboratory time
\begin{equation}\label{eq:Rpp1}
	R'' = - \frac{\langle rr'\rangle^2}{R^3} + \frac{\langle r'^2\rangle }{R} + \frac{\langle rr''\rangle}{R}\,.
\end{equation}
Substituting the harmonic approximation of the force from Eq. (\ref{eq:accel_ponderomotive}) in polar coordinates
\begin{equation}
	r'' = - \Omega^2 r + r\theta'^2
\end{equation}
into the expression for $R''$ provides
\begin{equation}\label{eq:Rpp}
	R'' + \left(\Omega^2 - \frac{\varepsilon_\perp^2}{\gamma^2 R^4}\right)R = 0\,,
\end{equation}
where
\begin{equation}
\varepsilon_\perp = \gamma \sqrt{\langle r^2 \rangle\langle r'^2 + r^2 \theta'^2\rangle - \langle rr'\rangle^2}
\end{equation}
is approximately the normalized transverse emittance of the bunch (see Appendix \ref{sec:emittance_def}). In the absence of energy spread and radiation reaction, $\varepsilon_\perp$ is a constant of motion: in a conservative potential, the phase-space distribution maintains a constant area despite deformations of its boundaries \cite{goldstein}. However, the initial ramp up and final ramp down of the pulse render the ponderomotive potential non-conservative and  change the emittance (see discussion in Section \ref{sec:radiation_cooling}).

Equation (\ref{eq:Rpp}) has an exact analytical solution for the initial condition $R(0) = R_0$ and $R'(0) = 0$:
\begin{equation}\label{eq:RMSt}
	R(t) = R_0 \sqrt{1 + \left(\frac{\varepsilon^2_\perp}{\gamma^2 R_0^4 \Omega^2} - 1 \right)\sin^2(\Omega t)}\,.
\end{equation}
In the absence of ponderomotive confinement, i.e., $\Omega \rightarrow 0$, Eq.(\ref{eq:RMSt}) demonstrates that the RMS radius increases without bound as $R(t) = R_0\sqrt{1+(t/T_s)^2}$, where $T_s = \gamma R_0^2/\varepsilon_\perp$. This expression describes the evolution of the RMS radius before the FF ramps up and after it ramps down. With ponderomotive confinement, the RMS radius either oscillates with the angular frequency $\Omega$ or remains constant. Setting the terms in the round brackets to zero provides the condition for constant RMS radius
\begin{equation}\label{eq:matching}
	\frac{\sigma_0\varepsilon_\perp}{\xi_0 R_0^2} = 1\,,
\end{equation}
where Eq. (\ref{eq:period}) has been used. Note that the dependence on the energy of the particles is still contained in the definition of the emittance. Although this formula applies only in the harmonic approximation, it provides a starting point for initializing the particle bunches in the simulations described in Section \ref{sec:simulations}.

\subsection{Longitudinal motion}\label{sec:longitudinal}
\begin{figure}
	\includegraphics[width=\linewidth]{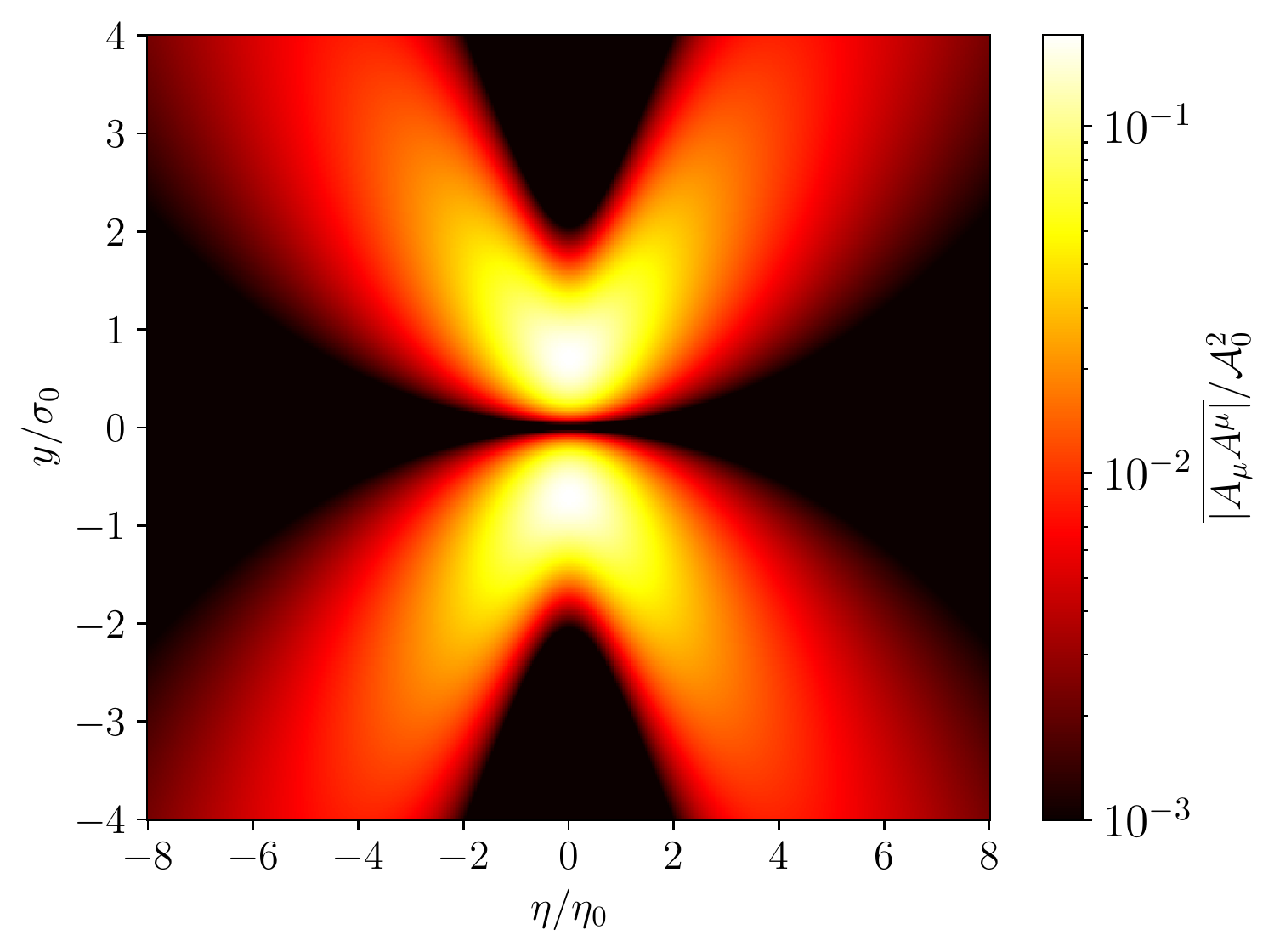}
	\caption{\label{fig:Along}Cycle-averaged invariant $|A_\mu A^\mu|$ for the $\ell=1$ OAM FF field in the plane of the field polarization. Note the different scales for the axes.}	
\end{figure}
Because the intensity peak of the FF pulse travels at the vacuum speed of light, the particles will gradually fall behind the peak intensity and experience weaker transverse confinement. There are several effects that can contribute to the rate at which the particles fall behind the peak. First, the initial velocity of each particle is less than the vacuum speed of light. This causes the lightcone variable $\eta = t + z$ to grow linearly in time, but this growth is negligible for ultra-relativistic particles. 

Second, the FF pulse can accelerate (or decelerate) the particles in the longitudinal direction. Fig. \ref{fig:Along} displays a slice of the $|A_\mu A^\mu|$ invariant in the polarization plane ($yz$-plane). In the focal region, this invariant varies weakly in the longitudinal direction, so that the ponderomotive force can be neglected. However, if the relativistic factor  $\gamma$ becomes comparable to the field strength $\xi_0$, the increase in the effective mass of the particles due to transverse and longitudinal oscillations in the fields of the FF pulse can significantly reduce the time-averaged longitudinal velocity. This deceleration also causes the lightcone variable to grow linearly in time and can be neglected as long as $\xi_0 \ll \gamma_0$. 

Finally, as was discussed in Ref. \cite{Formanek_2022}, a charged particle co-moving with the FF intensity peak continuously loses energy due to radiation reaction. The resulting deceleration becomes dominant in regions of high field intensity. When $\xi_0 \ll \gamma_0$, the amplitude of the fast transverse oscillations are small compared to the spot size of the pulse. As a result, each particle locally experiences a plane-wave-like field. Because the ultrarelativistic particles primarily move in the opposite direction of the phase fronts, the approximations $u_- = \gamma - u_z \approx 2 \gamma$ and $\phi \approx 2t$ can be employed. The electron energy loss due to Landau-Lifshitz radiation reaction in plane wave fields is then given by \cite{Di_Piazza_2008_a}
\begin{equation}\label{eq:gammat}
	\gamma(t)\approx \frac{\gamma_0}{1+\kappa(t)}\,,
\end{equation}
where
\begin{equation}\label{eq:decel}
	\kappa(t) = \frac{4}{3}\gamma_0 r_q \omega_0^2 \int_0^t \xi^2(t')dt'
\end{equation}
is the deceleration factor after a time $t$. The integral is taken over the normalized field amplitude $\xi(t')$ along the particle trajectory. This integral can be approximated by its average value  $\xi_\text{eff}^2 t$, where $\xi_\text{eff} < \xi_0$ is the effective field strength along the particle trajectory up to time $t$. The deceleration factor increases with the initial gamma factor and with the field strength along the particle trajectory. For the $\ell = 1$ OAM pulses of interest here, the field intensity is lowest on axis and rises with radial distance (up to $r_\text{max}$ for confined particles). As a result, the particles predominantly radiate in the regions around the turning points. 

For ultrarelativistic particles with small transverse velocity, the delay behind the intensity peak can be approximately evaluated as
\begin{equation}
	\begin{split}
	\eta_d(t)= \int_0^t &[1-\beta_z(\wt{t})]d\wt{t} \\
	&\approx \frac{1}{2} \int_0^t \left(\frac{1}{\gamma^2(\wt{t})} + \beta_\perp^2(\wt{t}) \right) d\wt{t}\,.
	\end{split}
\end{equation}
Substituting the expression for $\gamma(t)$ [Eq. (\ref{eq:gammat})] and approximating $\beta^2_\perp(t) \approx \beta^2_\perp(0) + \xi_\text{eff}^2/2\gamma^2(t)$, where the second term accounts for the transverse velocity of the rapid oscillations in the field, one obtains
\begin{equation}\label{eq:lag}
	\begin{split}
	\eta_d&(t) \approx \frac{1+\xi_\text{eff}^2/2}{2}\frac{1}{\gamma_0^2} + \frac{1}{2}\beta^2_\perp(0) t \\
	&+ \left(1+\frac{\xi_\text{eff}^2}{2}\right)\left(\frac{2}{3}\frac{r_q\omega_0^2\xi_\text{eff}^2}{\gamma_0}t^2 + \frac{8}{27}r_q^2 \omega_0^4 \xi_\text{eff}^4 t^3\right)\,.
	\end{split}	
\end{equation}
The terms on the first line are the contributions from the particle moving at a subluminal velocity with non-zero transverse component, while the terms on the second line are the contributions from radiation energy loss. In order to keep the particle close to the focus, $\eta_d(t)/\eta_0 \ll 1$ needs to be satisfied during the whole interaction. 
\subsection{Space-charge effects}\label{sec:coulomb_text}
To assess the impact of space-charge forces on the particle motion, consider a particle bunch with the charge density \cite{Tamburini:2019tzo}
\begin{equation}\label{eq:positiondistrib_text}
	\rho(r,z) =  qN \frac{1}{2\pi\sigma^2_r} e^{-\frac{r^2}{2\sigma^2_r}} \lambda_L(z)\,,
\end{equation}
where $N$ is the number of particles and $\sigma_r$ the bunch width. The longitudinal distribution
\begin{equation}\label{eq:lambda_text}
	\lambda_L(z) = \frac{1}{2L}\left[\text{erf}\left(\frac{L-2z}{2\sqrt{2}\sigma_r}\right) + \text{erf}\left(\frac{L+2z}{2\sqrt{2}\sigma_r}\right) \right]
\end{equation}
is parameterized by the length scale $L$ and was chosen because it permits an analytical solution for the field \cite{Tamburini:2019tzo}. Comparing the strength of the ponderomotive force to the repulsive fields of the particle bunch provides a condition for when space-charge effects can be neglected (see Appendix \ref{sec:coulomb}):
\begin{equation}\label{eq:Ncondition}
	\begin{split}
	N \ll N_\text{sc} &= 0.21 \frac{4\pi m}{q^2_e} \frac{L\sigma_r}{\sigma_0} \xi_0^2 \\
	&= 8 \times 10^7 [\mu\text{m}]^{-1} \frac{L\sigma_r}{\sigma_0} \xi_0^2\,,
	\end{split}
\end{equation}
where the numerical value is given for electrons. As an example, a typical electron bunch from laser wakefield acceleration (LWFA) has a $\sim$pC of charge \cite{vieira2012influence}, $L = 7\lambda_0$, and $\sigma_r = 3 \lambda_0/2\sqrt{2}$ (the same parameters used in the simulations presented below, see Sections \ref{sec:parameters} and \ref{sec:bunch_init}). For a FF pulse with $\xi_0 = 10$ and $\sigma_0 = 3\lambda_0$, $N_\text{sc}  = 2 \times 10^{10} \gg N$, thus space-charge forces are negligible. Note that the space-charge repulsion would be even less important in a mixed species electron-positron bunch \cite{sarri2015generation}.
\section{Required pulse energy: FF pulses vs other schemes}\label{sec:energy_estimates}
A FF laser pulse requires substantially less energy than a conventional LG10 laser pulse to confine a relativistic particle bunch. For a conventional laser pulse, the interaction time is limited by the Rayleigh range. Extending the interaction time requires increasing the Rayleigh range and the focal spot, which, in turn, requires increasing the power to maintain the strength of the ponderomotive force. In contrast, the intensity peak of a FF pulse co-propagates with the electron bunch, which decouples the interaction time from the Rayleigh range and the strength of the ponderomotive force.

Using Eq. (\ref{eq:totenergy}), the energy required in a FF pulse for the interaction time $t_\text{int,F}$ is given by
\begin{equation}\label{eq:EFF}
	E_\text{F} = 2P_\text{ave} t_\text{int,F} = \frac{\pi}{2}\mathcal{A}_{0,\text{F}}^2 \omega_0^2 \sigma_{0,\text{F}}^2 t_\text{int,F}\,,
\end{equation}
where the subscript F denotes the parameters of the FF pulse. Similarly, for a conventional LG10 pulse, denoted by subscript C, the energy is
\begin{equation}\label{eq:EG}
	E_\text{C} = \frac{\pi}{2}\mathcal{A}_{0,\text{C}}^2 \omega_0^2 \sigma_{0,\text{C}}^2 t_\text{int,C}\,.
\end{equation}
The interaction length in both cases is equal to the interaction time $L_\text{int} = t_\text{int}$. Confinement of the relativistic particles depends on the strength of the ponderomotive force. For a fixed ponderomotive force at a given distance from the axis [see Eq. (\ref{eq:accel_ponderomotive})]
\begin{equation}\label{eq:ponderomotive_ratio}
	\frac{\mathcal{A}_{0,\text{F}}^2}{\sigma_{0,\text{F}}^2} = \frac{\mathcal{A}_{0,\text{C}}^2}{\sigma_{0,\text{C}}^2} = K\,,
\end{equation}
where $K$ is proportional to the strength of the ponderomotive force. Substituting Eq. (\ref{eq:ponderomotive_ratio}) into Eqs. (\ref{eq:EFF}) and (\ref{eq:EG}) yields
\begin{align}
	\label{eq:EF}E_\text{F} &= \frac{\pi}{2} K \omega_0^2 \sigma_{0,\text{F}}^4 t_\text{int,F}\,,\\
	\label{eq:EC}E_\text{C} &= \frac{\pi}{2} K \omega_0^2 \sigma_{0,\text{C}}^4 t_\text{int,C}\,.
\end{align}
To ensure that the particles interact with the focus of the conventional pulse over the entire interaction time (or length)
\begin{equation}\label{eq:gaussian_range}
	t_\text{int,C} = 2\eta_{0,\text{C}} = \omega_0 \sigma_{0,\text{C}}^2\,.
\end{equation}
Thus, the energy in the conventional pulse grows as the cube of the interaction time [Eq. (\ref{eq:EC})], while the energy in the FF pulse scales linearly with the interaction time [Eq. (\ref{eq:EF})]. Now, two comparisons can be made:
\begin{itemize}
	\item[a)] For the same interaction time $t_\text{int,F} = t_\text{int,C} = t_\text{int}$
	\begin{equation}
		\frac{E_\text{C}}{E_\text{F}} = \frac{\sigma_{0,\text{C}}^4}{\sigma_{0,\text{F}}^4} =  \frac{t_\text{int}^2}{\omega_0^2 \sigma_{0,\text{F}}^4} = \left(\frac{t_\text{int}}{\eta_{0,\text{F}}}\right)^2\,.
	\end{equation}
    As an example, to confine an electron bunch with a radius of 2 $\mu$m over an interaction distance $L_\text{int}$ = 6 mm ($t_\text{int}$ = 20 ps), $E_\text{C}/E_\text{F} = 1.1\times 10^4$ where $\eta_{0,\text{F}} = 18 \pi\ \mu$m was used. Setting $\xi_0$ = 5, $E_\text{F}$ = 200 J and $E_\text{C} \approx $ 2 MJ. 
	\item[b)] For the same energy $E_\text{F} = E_\text{C}$
	\begin{equation}
		t_\text{int,F} = \left(\frac{t_\text{int,C}}{\eta_{0,\text{F}}}\right)^2 t_\text{int,C}\,.
	\end{equation}
    Thus, a FF pulse is advantageous as long as the interaction time $t_\text{C}$ is longer than the Rayleigh range of the FF pulse. An electron bunch with a radius of 2 $\mu$m can be confined by a FF laser pulse with $\eta_{0,\text{F}} = 18\pi\ \mu$m for a distance $L_\text{int,F}$ = 6 mm ($t_\text{int,F}$ = 20 ps) compared to only $L_\text{int,C}$ = 0.3 mm  ($t_\text{int,C}$ = 0.9 ps) for a conventional pulse, where both pulses have 200 J of energy.
\end{itemize}

Recently an alternative scheme that employs a Bessel beam to guide a relativistic electron bunch has been proposed \cite{Schachter:2020jaw,schachter2022normalized}. In this scheme, the electrons counter-propagate with respect to a radially polarized Bessel beam created by an axicon lens. The axicon creates an extended longitudinal region of high intensity. However, to create a ponderomotive barrier of comparable strength \cite{durnin1988comparison} to the FF, the axicon must maintain a high intensity across the entire region for the full interaction time. Maintaining this high intensity requires very high energies. In contrast, FF pulses concentrate the energy density along the the trajectory of the charged particles, greatly reducing the required energy.
\section{Simulations}\label{sec:simulations}
The motion of charged particles in the FF pulse was simulated using the classical Landau-Lifshitz equation of motion Eq. (\ref{eq:LL}). This equation accounts for the radiative energy losses from both the fast oscillations at the laser phase and the slow oscillations in the ponderomotive potential. Inspection of Eq. (\ref{eq:LL}) shows that the derivative term in the radiation-reaction force can be neglected \cite{Tamburini_2010} if the field in the instantaneous rest frame of the electron does not vary significantly over a classical electron radius. This is the case in all of the simulations presented here, and this term is ignored.

The simulations were performed for electrons (mass $m = m_e$, charge $q= q_e < 0$, classical radius $r_q = r_e$, and normalized field strength $\xi_0 = |q_e|\mathcal{A}_0 / m_e$). However, because the ponderomotive force applies equally for positively and negatively charged particles, all of the results also describe the motion of positrons. 
\subsection{Simulation parameters}\label{sec:parameters}

For the simulation results presented in this work, the distances are measured in the units of $k_0^{-1} = 2\pi/\lambda_0$ and the time in units $\omega_0^{-1} = 2\pi/\lambda_0$. In these units, the classical electron radius $r_e = 1.77 \times 10^{-8}\ k_0^{-1}$. 

Numerical integration of the electron equations of motion was performed using a fourth-order Runge-Kutta scheme with a time step $dt = 0.05\ \omega_0^{-1}$ and a total integration time $t_\text{int} = 4 \times 10^4\ \omega_0^{-1} = 21.2$ ps. A fifth order polynomial was employed to smoothly switch the fields on and off. The exact analytical form of the field envelope $g(\phi)$ can be found in the supplemental material of \cite{Formanek_2022}. The ramp time of the field was set to $\sim$ 0.4 ps and the period of the laser pulse was about 3.3 fs, corresponding to $\lambda_0 = 1 \mu \mathrm{m}$. Thus, the time scales were sufficiently disparate that the pulse envelope approximation and the expression for the laser power presented in Appendix \ref{sec:beam_power} were valid.

For all simulations the FF spot size was set to $\sigma_0 = 6\pi k_0^{-1} = 3\lambda_0$, which corresponds to the Rayleigh range $\eta_0 = 36\pi^2 k_0^{-1} = 18\pi \lambda_0$. Therefore the characteristic length in the longitudinal direction $\eta_0$ is about 19 times longer than the characteristic length in the transverse direction $\sigma_0$ (see discussion in Section \ref{sec:longitudinal}). The interaction length in terms of the Rayleigh range is $L_\text{int} = 112.6\ \eta_0$.

\subsection{Electron bunch initialization}\label{sec:bunch_init}
\begin{table}
	\begin{tabular}{c|ccc|c}
		\hline
		$\langle \gamma_0 \rangle$ & $\sigma_{\beta_\perp}(0)$ & $R^2(0)/k_0^{-2}$ & $\varepsilon_\perp(0)/\omega_0^{-1}$ & $\xi_0$ \\
		\hline \hline 
		1000 & 0.0021 & 44.4 & 13.9 & 5.9\\
		200 & 0.0054 & 45.2 & 7.10 & 3.0\\ 
		100 & 0.021 & 44.6 & 14.1 & 6.0\\
		\hline
	\end{tabular}
	\caption{\label{tab:parameters}Initial parameters for the electron bunches and normalized laser field amplitude $\xi_0$. The parameter $\xi_0$ was fixed so that the initial bunch satisfies the matching condition from Eq. (\ref{eq:matching}).}
\end{table}
\begin{figure*}
	\subfigure{\includegraphics[width=0.45\linewidth]{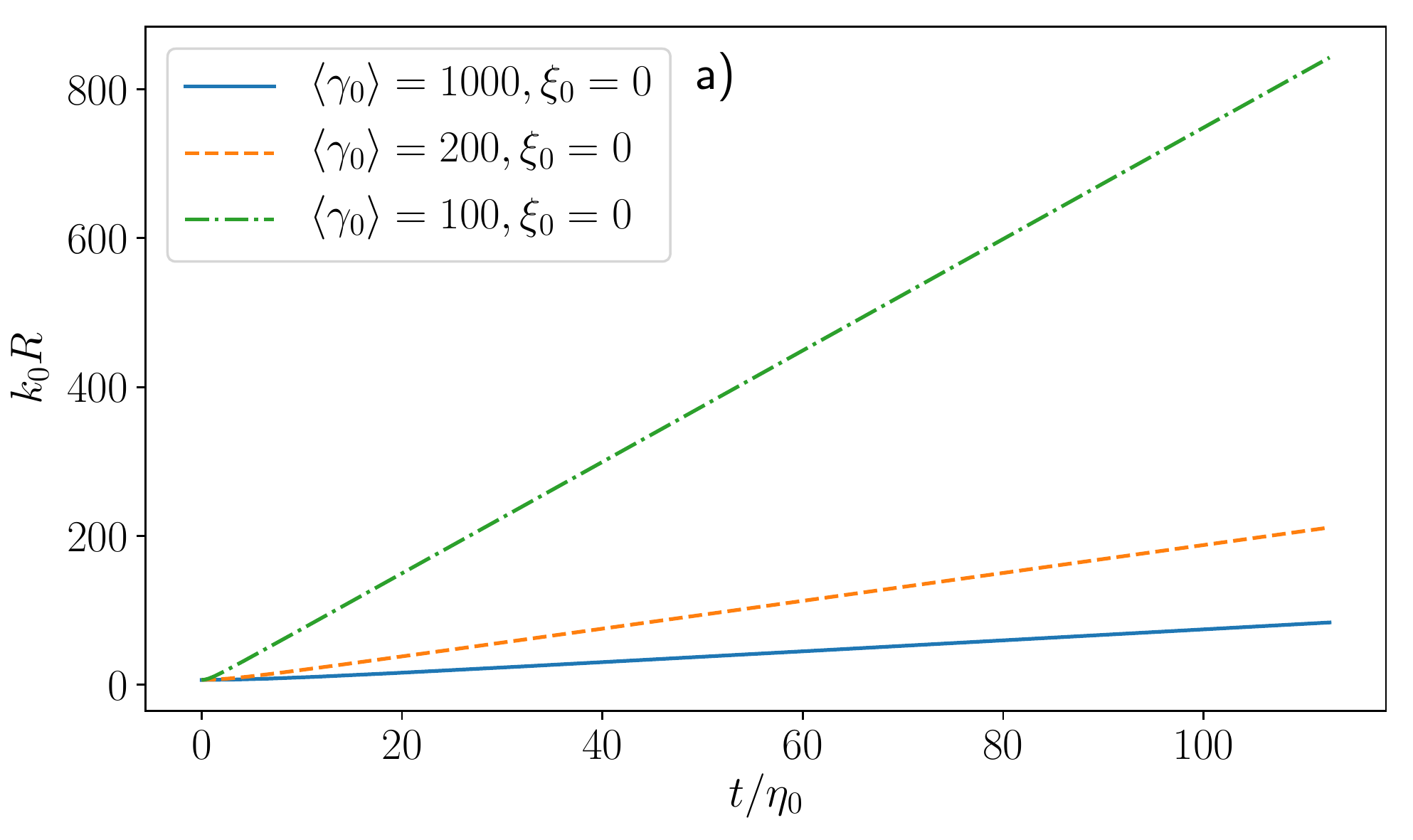}\label{fig:rmsr_t_a}}
	\subfigure{\includegraphics[width=0.45\linewidth]{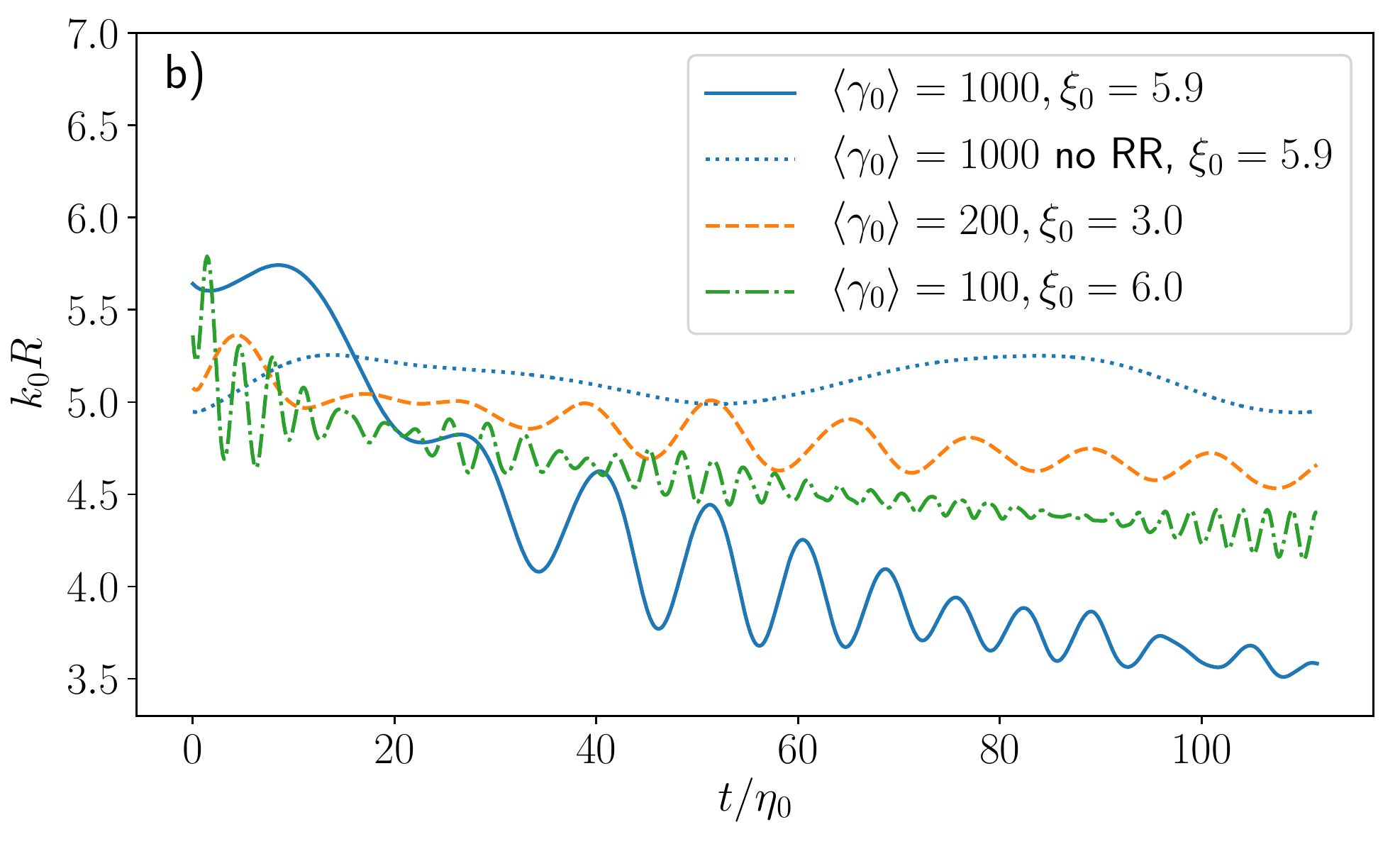}\label{fig:rmsr_t_b}}
	\caption{\label{fig:rmsr_t} Time evolution of the RMS radius for a) freely traveling electrons with no external field and b) electrons confined to the ponderomotive potential of a FF pulse with $\sigma_0 = 6\pi k_0^{-1} = 3\lambda_0$. The dotted line denotes the case of $\langle \gamma_0 \rangle = 1000$ with no radiation reaction. For the parameters of the electron bunches see Section \ref{sec:bunch_init}.}
\end{figure*}
The electron bunches were initialized based on parameters typical of LWFA with bunch lengths of $\sim 10\lambda_0$ and divergences up to tens of mrad \cite{vieira2012influence,sears2010emittance}. The electrons move predominantly in the negative $z$ direction with the intensity peak of the FF pulse and against the phase fronts (see Fig. \ref{fig:graphics}). The initial electron positions were randomly sampled from the charge distribution given by Eqs. (\ref{eq:positiondistrib_text}) and (\ref{eq:lambda_text}), with a width and length representative of bunches produced in either laser wakefield accelerators or proposed conventional accelerators \cite{Vranic:2018liw,ParticleDataGroup:2016lqr}. The electron bunch was aligned with the optical axis of the FF pulse, such that $\langle \bm{x}_\perp (0)\rangle = 0$. More generally, the centroid of the bunch would also oscillate in the ponderomotive potential [see Eq. (\ref{eq:centroid})]. The initial variance in the radial position was chosen to be 
\begin{equation}\label{eq:sigmar}
	\sigma_r(0) = \frac{r_\text{max}}{2} = \frac{\sigma_0}{2\sqrt{2}} = \frac{3\pi}{\sqrt{2}}  k_0^{-1} = \frac{3}{2\sqrt{2}}\lambda_0\,.
\end{equation}
The initial longitudinal spread of the bunch was set to
\begin{equation}\label{eq:L}
	L(0) = 14 \pi k_0^{-1} = 7 \lambda_0\,.
\end{equation}
With this choice, $\sim$99\% of the simulated electrons are initialized within a longitudinal distance of 5$\lambda_0$ from the center of the bunch. The length of the electron bunch is therefore much shorter than the Rayleigh range $18\pi \lambda_0$. As a result, electrons initialized within a longitudinal distance of $5\lambda_0$ from the focus experience a ponderomotive force that is within 99\% of the maximum.

The initial longitudinal components of the four-velocity were normally distributed with a standard deviation equal to 1\% of the central value $u_z(0) = -\sqrt{\langle \gamma_0 \rangle ^2 - 1}$. The gamma factors $\langle \gamma_0 \rangle$ used to generate the distribution were 1000, 200, and 100 for the three simulated cases. The transverse velocities were also normally distributed but with zero mean. The initial variance in the magnitude of perpendicular velocity $\sigma_{\beta_\perp}(0)$ and the normalized field strength $\xi_0$ were chosen such that the condition in Eq. (\ref{eq:matching}) was satisfied. See Table \ref{tab:parameters} for details. To capture the effect of electrons escaping the FF pulse, the initial values were also chosen to ensure that some electrons were initialized with a transverse velocity and position outside of the constraint in Eq. (\ref{eq:phasespace}). As a result, the subset of confined electrons were not normally distributed. With the longitudinal and transverse velocities known, the initial energy for each electron was fully determined. Each simulated bunch was composed of 1000 independent electrons, which was sufficient for calculating average quantities. 

For the simulation parameters given in Table \ref{tab:parameters}, the maximum quantum nonlinearity parameter for electrons $\chi_e = |q_e|\sqrt{|(F_{\mu\nu}u^\nu)^2|}/m^2$ (see Ref. \cite{Di_Piazza_2008_a}) is 0.03, 0.003, and 0.003, respectively. In addition, the De Broglie wavelengths of the simulated electrons were between 50 - 170 fm, which is many orders of magnitude smaller than the 1 $\mu$m laser wavelength. Thus, the use of classical radiation reaction is justified.
\subsection{Electron confinement in the FF pulse and radiation cooling}\label{sec:radiation_cooling}
\begin{figure}
	\includegraphics[width=\linewidth]{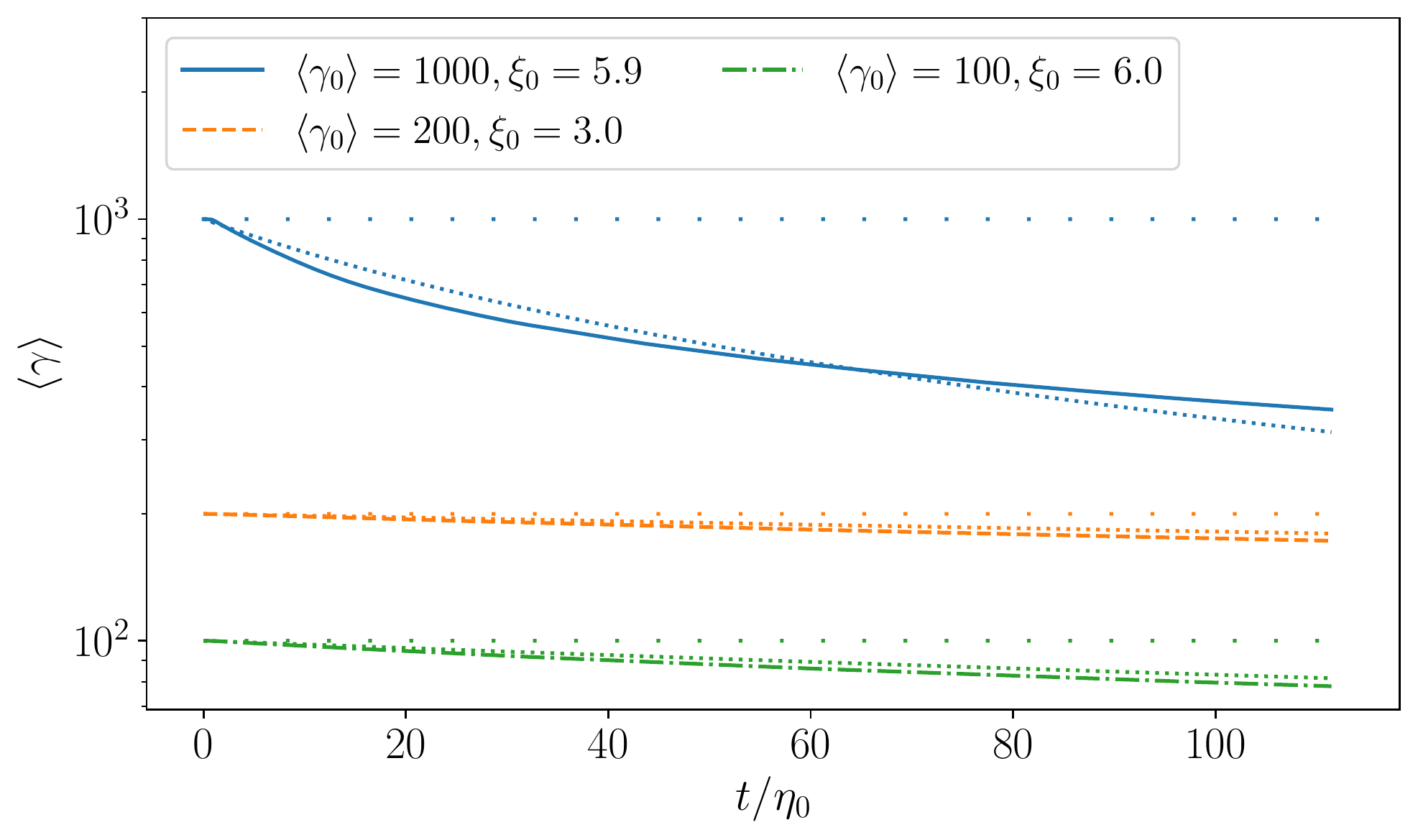}
	\caption{\label{fig:gamma_RR}Average relativistic gamma factor of electrons confined to the ponderomotive potential of the FF pulse with $\sigma_0 = 6\pi k_0^{-1} = 3\lambda_0$. The loosely dotted lines denote simulation runs with no radiation reaction. The densely dotted lines denote analytical estimates based on Eq. \ref{eq:gammat} with $\xi_\text{eff} = 0.26\ \xi_0$. For the parameters of the electron bunches see Section \ref{sec:bunch_init}.}
\end{figure}
Figure \ref{fig:rmsr_t_a} demonstrates the rapid expansion of an electron bunch in the absence of external fields. Early in time the expansion is slower as a subset of electrons move towards the bunch axis. Quickly thereafter, the initial spread in transverse momenta cause the RMS radius to evolve as approximately [see discussion below Eq. (\ref{eq:RMSt})]
\begin{equation}\label{eq:nofield}
	\left. R(t)\right|_{\xi_0 = 0} \approx R_0 \frac{t}{T_s} \approx \sigma_{\beta_\perp}(0) t\,.
\end{equation}
The beam divergences without the fields of the FF pulse
\begin{equation}
\begin{split}
\Theta = 2\arctan \left( \frac{R(t_\text{int}) - R(0)}{\langle\beta_z(0)\rangle t_\text{int}} \right) \\
\approx 2 \arctan \left(\frac{\sigma_{\beta_\perp}(0)}{\langle \beta_z(0)\rangle }\right)
\end{split}
\end{equation}
were 4.2, 11 and 42 mrad respectively. The electrons had a high enough transverse momentum so that some escaped the ponderomotive barrier of the FF pulse and also large enough beam divergences $> 1$ mrad to be relevant to LWFA-based electron sources \cite{sears2010emittance}.

As shown in Fig. \ref{fig:rmsr_t_b}, this expansion is contained by counter-propagating a flying focus pulse with the electron bunch. Even though the matching condition for a constant bunch radius, i.e., Eq. (\ref{eq:matching}), was satisfied for the mean energy, the energy spread in the bunch and the anharmonicity of the ponderomotive potential result in oscillations of the RMS radius. Only confined electrons, defined as those chosen as having $r(t) < 0.75 \sigma_0$ during the entire interaction, were used to calculate the RMS radius in Fig. \ref{fig:rmsr_t_b}. 

Radiation reaction gradually decreases the RMS spot size and increases the oscillation frequency of the bunch [cf. $\langle \gamma_0 \rangle$ = 1000 cases in Fig. \ref{fig:rmsr_t_b}]. As the electrons radiate and lose energy (Fig. \ref{fig:gamma_RR}), the ponderomotive force becomes stronger [Eq. (\ref{eq:accel_ponderomotive})], which increases the oscillation frequency [Eq. (\ref{eq:period})]. Consistent with Eq. (\ref{eq:gammat}), the radiative cooling of the bunch occurs more rapidly for higher values of $\gamma_0$ and $\xi_0$ (Fig. \ref{fig:gamma_RR}).

\begin{figure*}
	\subfigure{\includegraphics[width=0.45\linewidth]{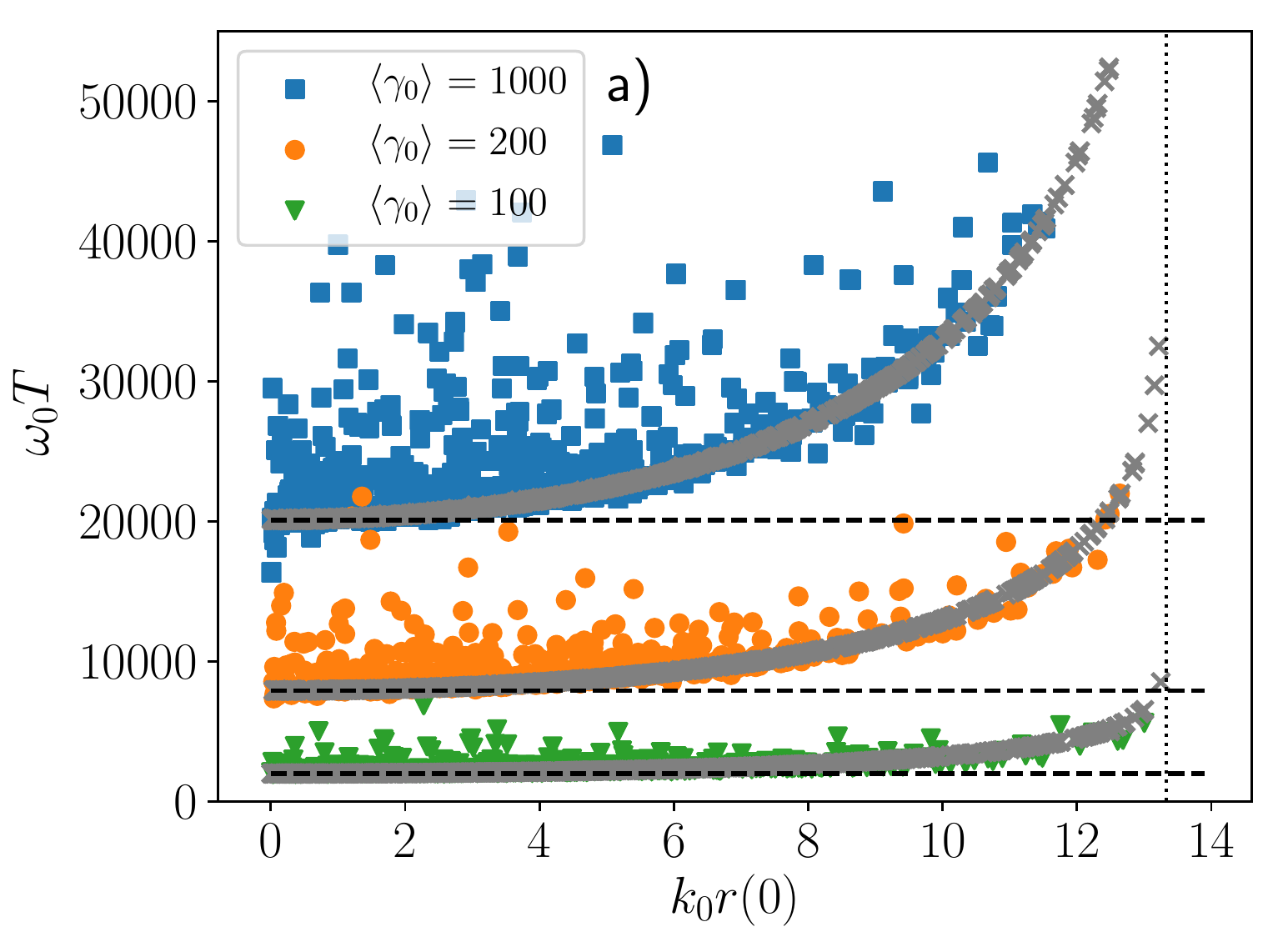}\label{fig:T_vs_r_a}}
	\subfigure{\includegraphics[width=0.45\linewidth]{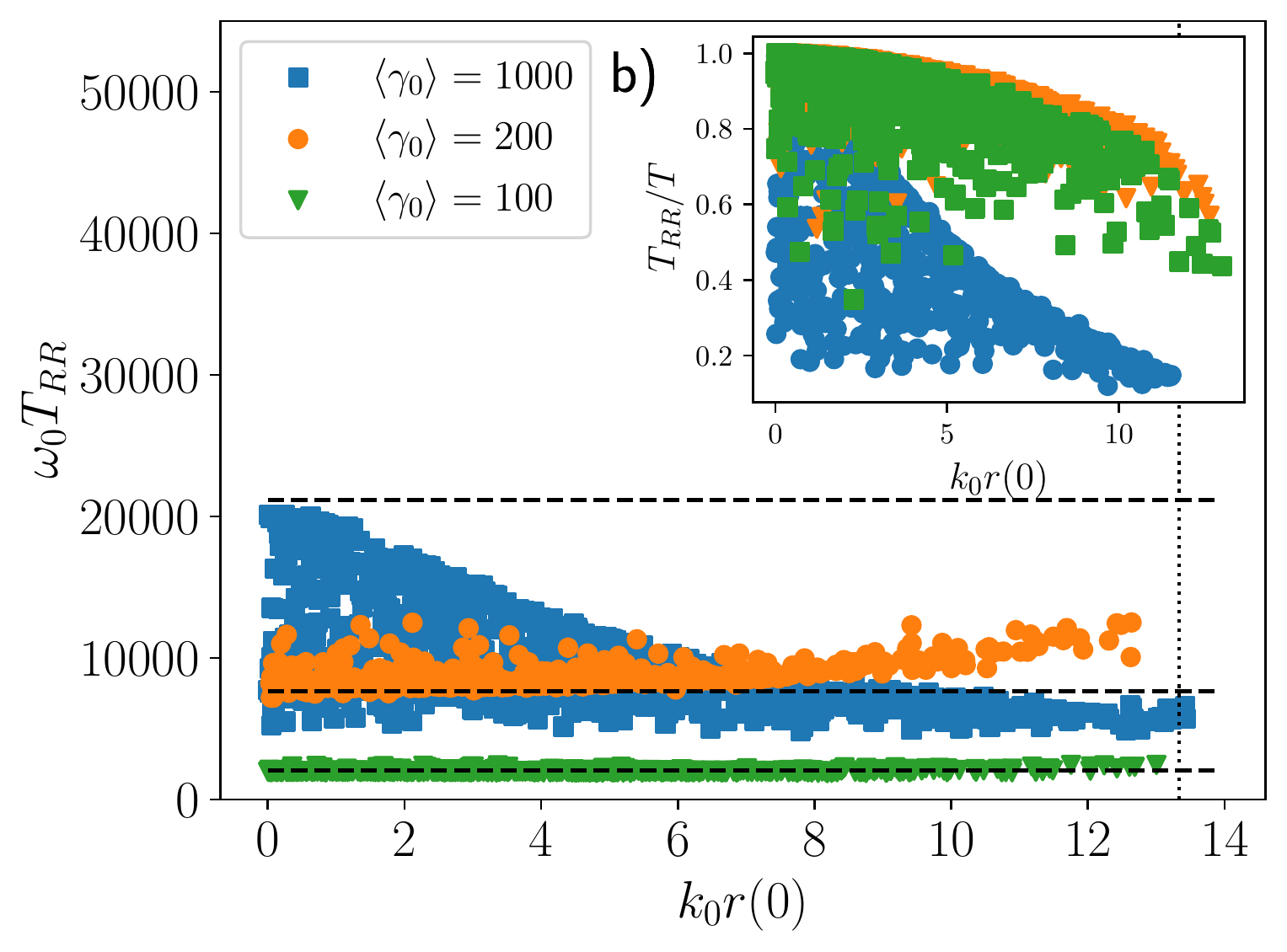}\label{fig:T_vs_r_b}}
	\caption{\label{fig:T_vs_r} Oscillation period in the ponderomotive potential as a function of the initial distance from the $z$-axis. The horizontal dashed lines indicate the period in the harmonic approximation, Eq. (\ref{eq:period}). The vertical dotted line marks $r_\text{max} = \sigma_0/\sqrt{2}$. a) radiation reaction is switched off. The gray crosses show the equivalent simulation runs with zero energy and momentum spread. b) radiation reaction is included. The inset displays the ratio of oscillation periods with and without radiation reaction. For the parameters of the electron bunches see Section \ref{sec:bunch_init}.}
\end{figure*}
The reduction in the RMS spot size of the bunch and increase in its oscillation frequency due to radiation reaction mitigate the effect of anharmonicity. Figure \ref{fig:T_vs_r} displays the oscillation periods of electrons as a function of initial radius without [Fig. \ref{fig:T_vs_r_a}] and with [Fig. \ref{fig:T_vs_r_b}] radiation reaction. The period of oscillations around an axis $i$ (either $x$ or $y$)  was determined numerically as an average period over the interaction time
\begin{equation}
	T_i = \frac{2(t^{(n_i)}_i-t^{(1)}_i)}{n_i - 1}\,,
\end{equation} 
where $n_i$ is the number of times the electron crosses the $i^\text{th}$ axis. The first crossing happens at time $t^{(1)}_i$ and last at time $t^{(n_i)}_i$. The arithmetic mean of $T_x$ and $T_y$ is plotted in Fig. \ref{fig:T_vs_r}.

Without radiation reaction [Fig. \ref{fig:T_vs_r_a}], electrons initialized at small radii oscillate with a period close to that predicted by Eq. (\ref{eq:period}), marked by the horizontal dashed lines.  In contrast, electrons initialized at larger radii undergo oscillations with a longer period due to the weakening of the ponderomotive potential with increasing radius. Figure \ref{fig:T_vs_r_b} and its inset demonstrate the decrease in the oscillation period resulting from radiation reaction as compared to Fig. \ref{fig:T_vs_r_a}. The decrease in period is most pronounced for electrons initialized further from the $z$-axis in regions of high intensity, where radiation reaction is strongest. The effect of radiation reaction diminishes with decreasing $\gamma_0$ and $\xi_0$ as predicted by Eq. \ref{eq:gammat}.

\begin{figure}
	\subfigure{\includegraphics[width=\linewidth]{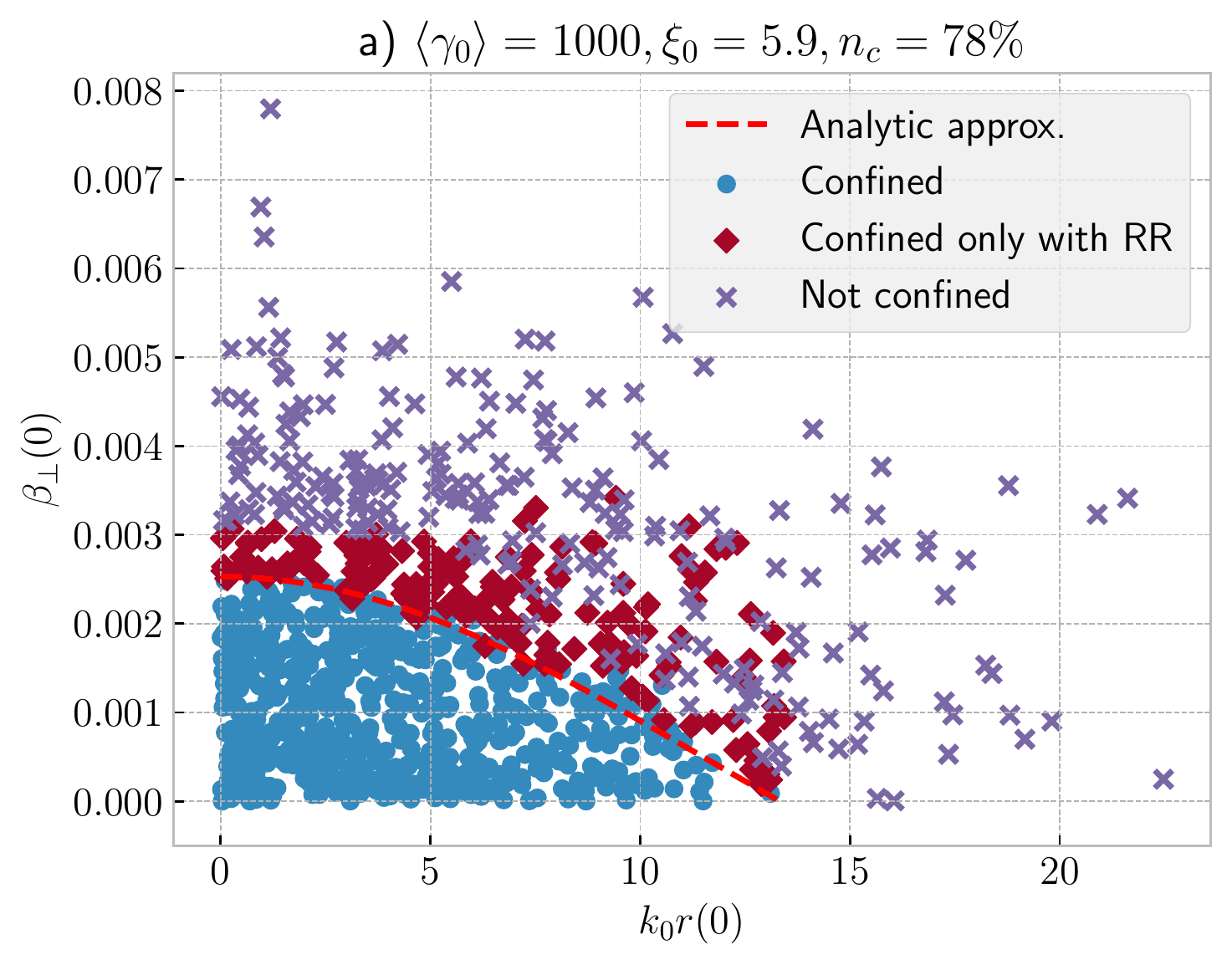}\label{fig:phase_diagram_a}}\\
	\subfigure{\includegraphics[width=\linewidth]{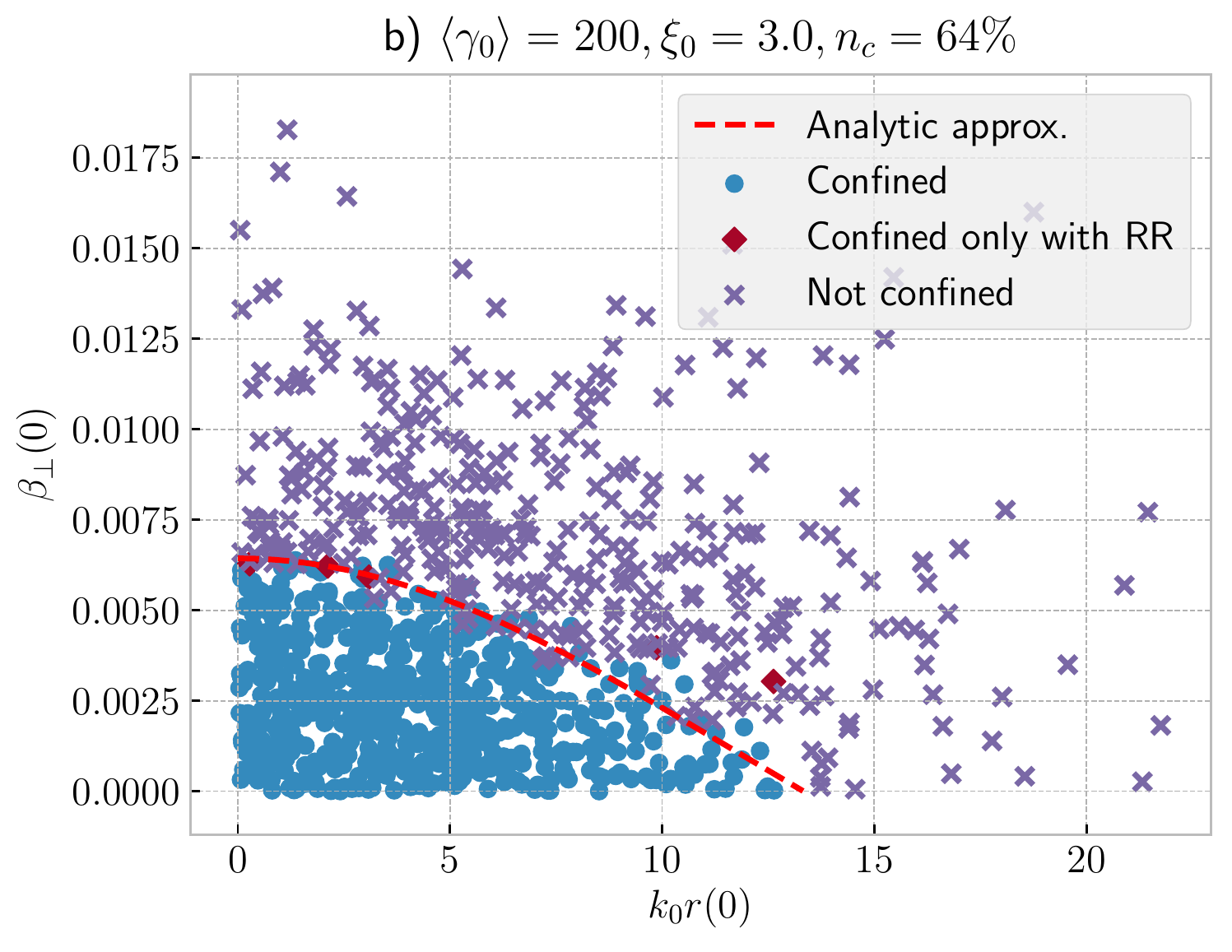}\label{fig:phase_diagram_b}}\\
	\subfigure{\includegraphics[width=\linewidth]{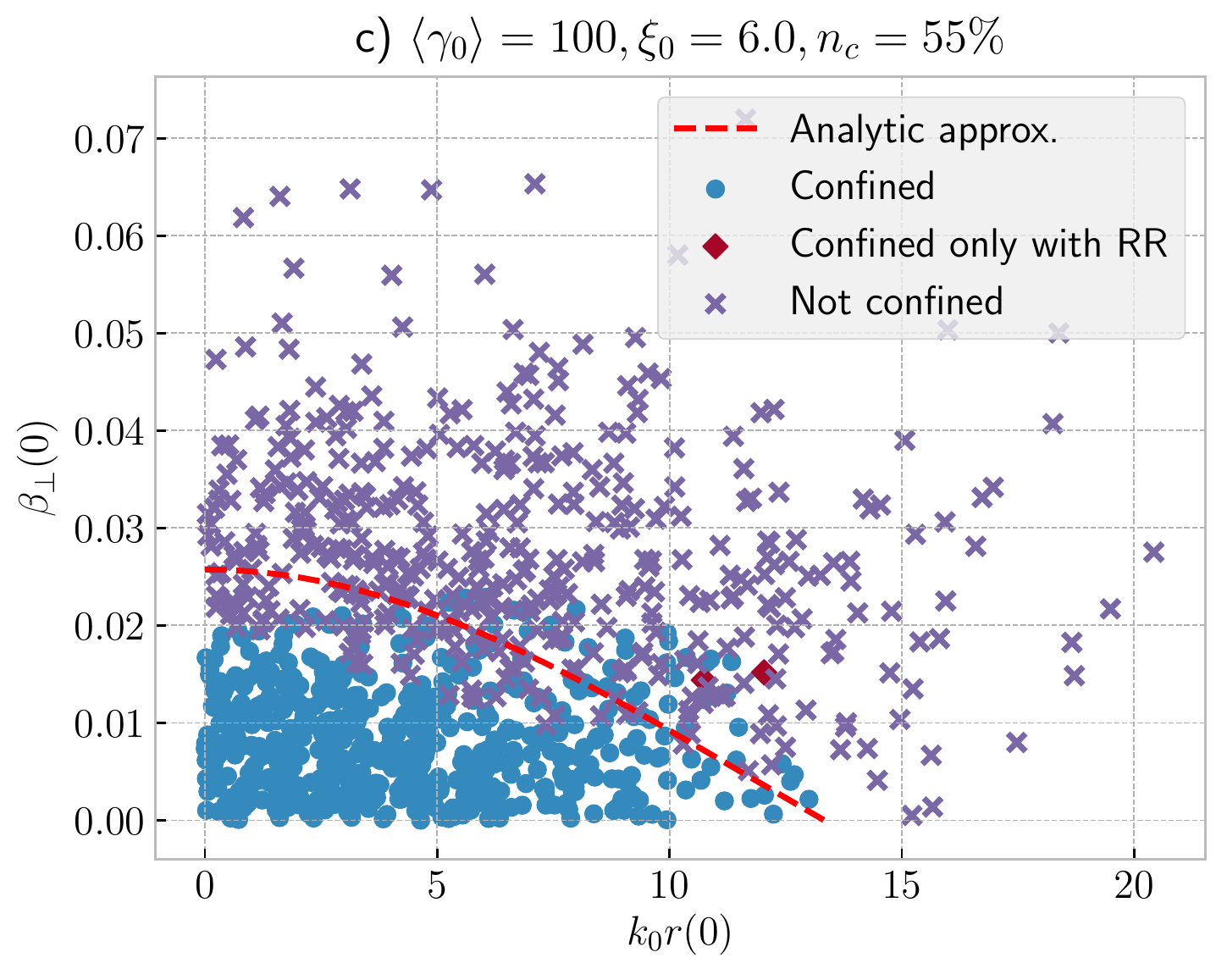}\label{fig:phase_diagram_c}}
	\caption{\label{fig:phase_diagram}Initial transverse phase space of electrons co-travelling with the intensity peak of a FF pulse, with $\sigma_0 = 6\pi k_0^{-1} = 3\lambda_0$. The analytical approximation for the boundary between confined and not confined electrons (red dashed line) is given by Eq. (\ref{eq:phasespace}). Each simulation evolved 1000 independent electron trajectories and $n_c$ indicates the percentage of confined electrons. For the parameters of the electron bunches see Section \ref{sec:bunch_init}.}
\end{figure}
The increase in the strength of the ponderomotive potential as electrons lose energy to radiation reaction results in the confinement of more electrons. This is illustrated in Fig. \ref{fig:phase_diagram_a} which shows the initial phase space distributions of confined and unconfined electrons for $\langle \gamma_0 \rangle = 1000 $. The trapping boundary predicted by Eq. (\ref{eq:phasespace}) is also plotted as a red dashed line. Consistent with the reduction in period shown in Fig. \ref{fig:T_vs_r}, the increase in trapping is most pronounced for electrons initialized in regions of high intensity [electrons well outside of the red dashed line in Fig. \ref{fig:phase_diagram_a} are still confined]. 

In Figs. \ref{fig:phase_diagram_b}  and \ref{fig:phase_diagram_c}, $\langle \gamma_0 \rangle$ is lower, which reduces the effect of radiation reaction on the electron trajectories. The highest $\xi_0 / \gamma_0$ ratio is presented in Fig. \ref{fig:phase_diagram_c}. In this case the coupling between longitudinal and transverse motion becomes important, and Eq. (\ref{eq:phasespace}) is no longer accurate, which can be observed as the lack of confinement within the red-dashed boundary.

\begin{figure}
	\includegraphics[width=\linewidth]{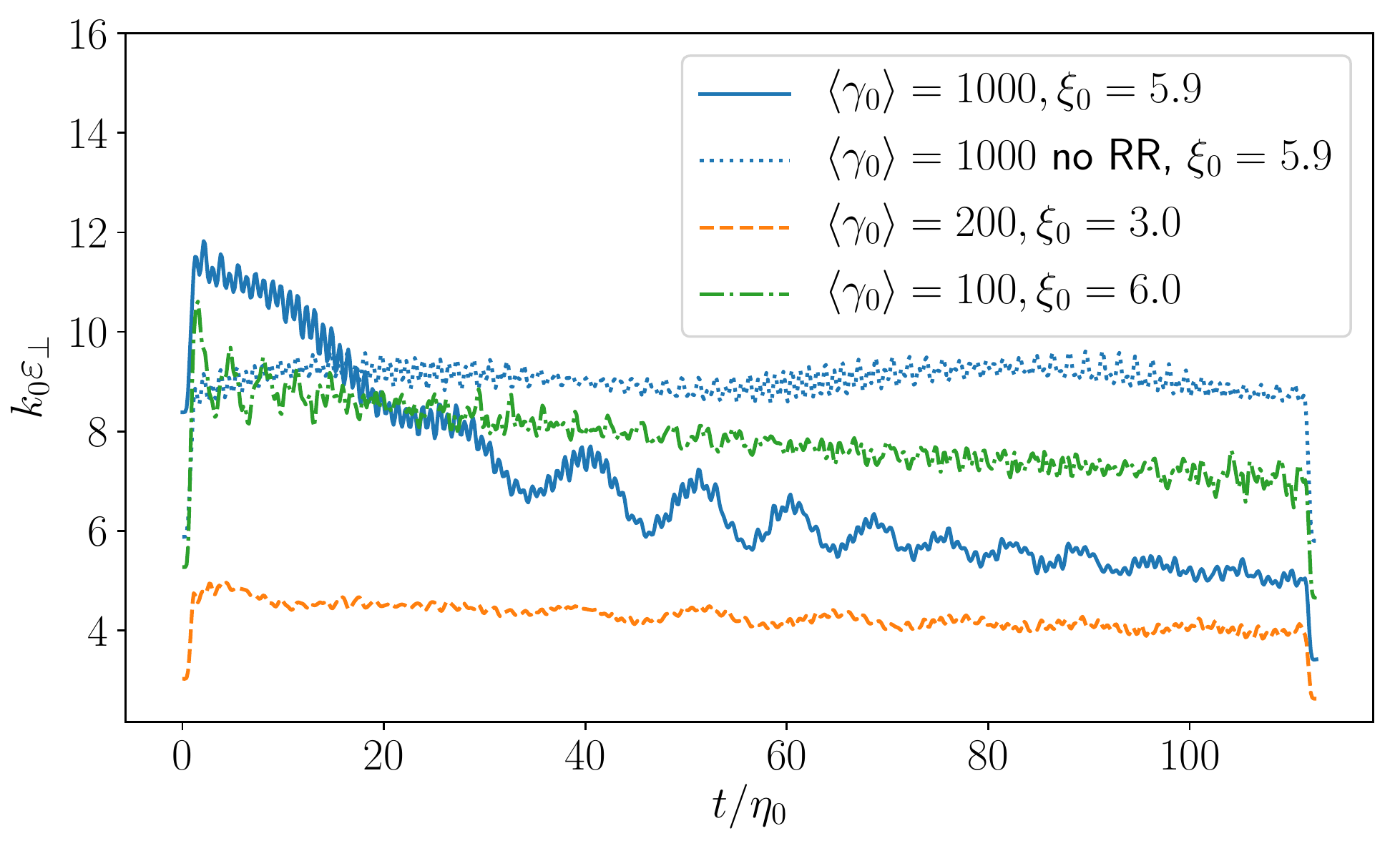}
	\caption{\label{fig:emittance_RR} Time evolution of the normalized transverse emittance of confined electrons co-traveling with the intensity peak of a FF pulse, with $\sigma_0 = 6\pi k_0^{-1} = 3 \lambda_0$. The $\langle \gamma_0 \rangle = 1000$ case with no radiation reaction is shown as the dotted line. For the parameters of the electron bunches see Section \ref{sec:bunch_init}. }
\end{figure}
\begin{table}
	\begin{tabular}{c|cc}
		\hline
		$\langle \gamma_0\rangle $ & $\Delta \langle \gamma \rangle$ 	&	$\Delta  \varepsilon_\perp $\\
		\hline \hline
		1000 & 0.65	&	0.55\\
		200 & 0.14	&	0.17\\
		100 & 0.22	&	0.24\\
		\hline
	\end{tabular}
	\caption{\label{tab:rel_change}Relative changes in the average gamma factor from Fig. \ref{fig:gamma_RR} and in the normalized transverse emittance from Fig. \ref{fig:emittance_RR} for the three simulated cases.}
\end{table}

In addition to reducing the RMS radius of the electron bunch and improving the transverse confinement, radiation reaction reduces the emittance of the electron beam (Fig. \ref{fig:emittance_RR}). However, all of these improvements in the quality of the electron bunch come at the expense of its average energy (Fig. \ref{fig:gamma_RR}). In fact, comparing the $\langle \gamma_0 \rangle = 1000$ cases in Figs. \ref{fig:gamma_RR} and \ref{fig:emittance_RR} shows that the relative change in the emittance $\Delta \varepsilon_\perp = [\varepsilon_\perp(0) - \varepsilon_\perp(t_\text{int})] / \varepsilon_\perp(0)$ is nearly equal to the relative change in the average energy $\Delta \langle \gamma \rangle = [\langle \gamma_0 \rangle - \langle \gamma(t_\text{int})\rangle ]/\langle \gamma_0 \rangle$: 
\begin{equation}
	\Delta \varepsilon_\perp \approx \Delta \langle \gamma \rangle\,.
\end{equation}
For example, in the $\langle \gamma_0 \rangle = 1000$ case, the average gamma factor decreases by 65\% from 1000 to 353 while the emittance drops by 55\% from 11.2 $k_0^{-1}$ to 5.0 $k_0^{-1}$. Due to the oscillatory nature of the emittance, an average value from $5\,\eta_0$ after the field ramp up and from $5\,\eta_0$ before the field ramp down was taken. The relative changes for the remaining two electron bunches are shown in Table \ref{tab:rel_change}.

In Fig. \ref{fig:emittance_RR}, the emittance was calculated using the electrons that remained confined to the FF pulse, i.e., those with $r(t) < 0.75\sigma_0$ during the entire interaction. The difference in the initial emittances of the beams with and without radiation in Fig. \ref{fig:emittance_RR} was due to the different statistics of the confined electrons in the two cases. The jump in emittance during the ramp on and ramp off of the FF pulse results from the onset of electron oscillations in the fields: the statistical definition of emittance [Eq. (\ref{eq:emittance})] uses mechanical and not canonical transverse momentum.
\subsection{Longitudinal motion}\label{sec:longitudinal_simulations}
\begin{figure}
	\includegraphics[width=\linewidth]{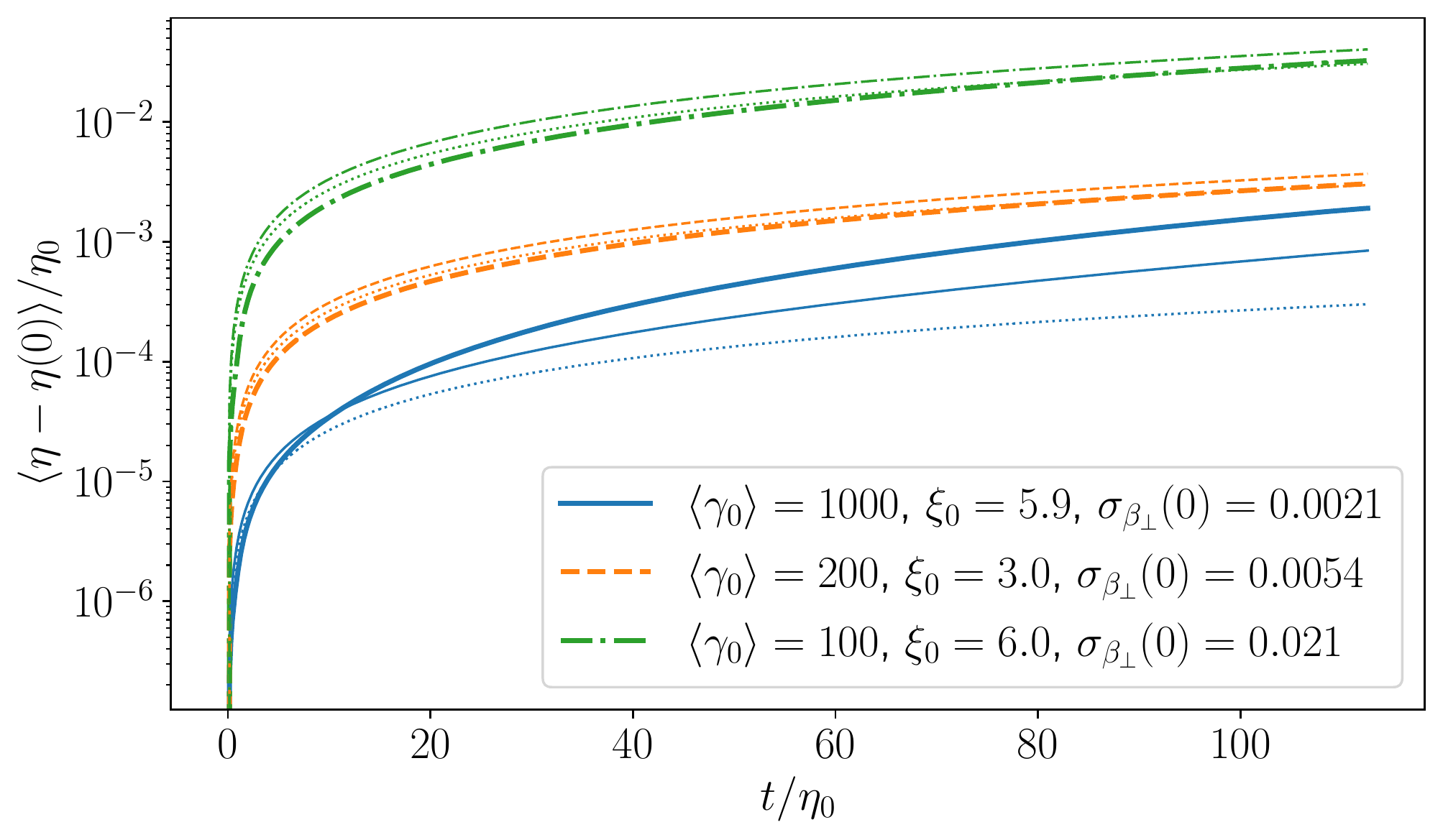}
	\caption{\label{fig:zspread_RR} Average longitudinal displacement of confined electrons from the focus of the FF pulse with $\sigma_0 = 6\pi k_0^{-1} = 3\lambda_0$ (thick lines). Thin lines indicate the estimate of Eq. (\ref{eq:lag}) for the longitudinal lag. The dotted lines show the results from simulations without the fields of the FF pulse. For the parameters of the electron bunches see Section \ref{sec:bunch_init}.}
\end{figure}
Over the entire interaction length, the electron bunch remains in the vicinity of the intensity peak to within a small fraction of the Rayleigh range (Fig. \ref{fig:zspread_RR}). The longitudinal delay of the bunch with respect to the FF intensity peak is in reasonable agreement with the predictions of Eq. \ref{eq:lag}. In Fig. \ref{fig:zspread_RR}, the average delay for the three simulated bunches are plotted as thick lines, while Eq. (\ref{eq:lag}) is plotted as thin lines. For the purposes of Fig. \ref{fig:zspread_RR}, Eq. (\ref{eq:lag}) was evaluated using average quantities of the bunch, i.e., $\beta^2_\perp(0) \rightarrow \sigma^2_{\beta_\perp}(0)$ and $\gamma_0 \rightarrow \langle \gamma_0 \rangle$, and the effective field strength along the particle trajectory was set to the fit value obtained from Fig. \ref{fig:gamma_RR}: $\xi_\text{eff} = 0.26\xi_0$. 

The increase in delay due to radiation reaction is only significant for the $\langle \gamma_0 \rangle = 1000$ case. For the $\langle \gamma_0 \rangle = $ 100 and 200 cases, the delay results almost entirely from the initial subluminal velocity of the electrons. This is demonstrated by the similarity of the thick lines and the dotted lines which show the average delay in the absence of the fields of the FF pulse.  

Any predictable delay can be eliminated by using a FF pulse with a focal velocity equal to that of the electrons. Closed form expressions for FF pulses with focal velocities $\beta_F \neq -1$ have been derived in the paraxial approximation \cite{Ramsey_2022} and exactly \cite{Ramsey_2022b}. However, this more general treatment is not necessary here, because the average electron delay $\langle \eta -\eta(0)\rangle$ is much smaller than the Rayleigh range $\eta_0$ over the entire interaction length. 
\section{Summary and conclusions}
A flying focus pulse with $\ell = 1$ OAM can prevent the spreading of relativistic particle bunches over macroscopic distances, providing an alternative to magnetic optics at high power laser facilities. The peak intensity of the FF pulse travels at the vacuum speed of light in the opposite direction as its phase fronts. Charged particles traveling with the peak intensity experience a ponderomotive potential that confines their transverse motion over distances far greater than a Rayleigh range. Radiation reaction decreases the RMS radius and emittance of the particle bunch and improves the transverse confinement at the cost of a reduction in the average particle energy. Simulations demonstrated the confinement of 50 - 500 MeV electron bunches with 40 - 4 mrad beam divergences over 6 mm. The electron bunches maintained a tight RMS radius of $ \sim1\, \mu \mathrm{m}$. 

All-optical confinement of a charged particle bunch with a FF could have utility in any situation where the bunch must be transported from its source to its target with a small RMS radius. For instance, the transverse size of the particle bunch determines the spatial resolution of probes based on secondary radiation sources, such as bremsstrahlung x-ray imaging. The bunch size also contributes to the Pierce parameter, which is critical to the performance of free-electron-lasers. 

Flying focus pulses require much less energy to confine a charged particle bunch than either LG10 Gaussian or axicon-focused Bessel pulses. In contrast to these pulses, the peak intensity of the FF travels with the electron bunch, which decouples the interaction length from the Rayleigh range. For the simulated examples, the FF pulse had an energy and power of ~200 J and $<5$ TW, respectively, compared to the 2 MJ required in a LG10 Gaussian pulse. 

The distance over which a particle bunch remains confined can be lengthened by using FF pulses with a peak intensity that travels at a velocity equal to that of the particles. Such pulses have been experimentally demonstrated and theoretically analyzed \cite{Froula_2018,Ramsey_2022b}. The use of a velocity-matched FF pulse would provide an additional advantage over Gaussian or Bessel pulses. 

The electron bunches considered in this work had parameters characteristic of the bunches created in LWFA. The short lengths ($\sim 10\, \mu$m) and high divergences ($> 1$ mrad) make these bunches ideal for confinement by a FF pulse. The short length also ensures that the bunch sits in a region of near-constant peak intensity. The high divergences ensure that the confinement afforded by the FF pulse has an impact on the transport. At high-intensity laser facilities, a laser pulse can be used for both LWFA and transport of the resulting bunches, without the need for magnetic optics. In contrast, conventional e-/e+ accelerators, such as those produced at SLAC, have much longer bunch lengths ($\sim 1$ mm) and smaller divergences ($< 1$ mrad). However, shorter bunch lengths are expected for next generation e-/e+ colliders, such as the ILC or CLIC. 
 
While the simulations were performed for electrons, the results are equally applicable to positrons. In fact, mixed electron-positron bunches \cite{sarri2015generation} experience less Coulomb repulsion due to their lower net charge and would be easier to confine. This property could be exploited to guide the products of Breit-Wheeler pair production from the collision of a high-intensity laser pulse with hard photons. Moreover, transverse confinement in a FF pulse could provide an alternative to injecting electron beams for mitigating alignment sensitivity in wakefield and direct laser acceleration of positrons \cite{lindstrom2018measurement,martinez2022creation}. 
\begin{acknowledgments}
	We would like to thank Konstantin Beyer, Dustin Froula, Yevgeny Gelfer, Ond\v{r}ej Klimo, Francesco Schillaci, Stefan Weber, and Kathleen Weichman (ordered alphabetically) for enlightening discussions and Levi Sch\"{a}chter for correspondence. The work of MV is supported by the Portuguese Science Foundation grant FCT No. CEECIND/01906/2018 and PTDC/FIS- PLA/3800/2021. The work of JPP and DR is supported by the Office of Fusion Energy Sciences under Award Number DE-SC0019135 and DE-SC00215057, the Department of Energy National Nuclear Security Administration under Award Number DE-NA0003856, the University of Rochester, and the New York State Energy Research and Development Authority. This publication is also supported by the Collaborative Research Centre 1225 funded by Deutsche Forschungsgemeinschaft (DFG, German Research Foundation)—Project-ID 273811115—SFB 1225. This project has received funding from the European Union’s Horizon Europe research and innovation programme under the Marie Sk\l{}odowska-Curie grant agreement No. 101105246.
\end{acknowledgments}

\appendix
\section{Electromagnetic fields of $\ell=1$ OAM FF beam}\label{sec:emfields}
Here, the electromagnetic field components of a FF beam are derived for the special case where the focal velocity is equal and opposite to the phase velocity at the vacuum speed of light. The vector four-potential $A^\mu$, linearly polarized along $y$-direction, is fully determined by Eqs. (\ref{eq:aperp}), (\ref{eq:aminus}), and (\ref{eq:A0}). The electromagnetic field components can be calculated using the standard formulas
\begin{equation}
	\bm{E} = -\partial_t \bm{A} - \bm{\nabla} A^0, \quad \bm{B} = \bm{\nabla} \times \bm{A},
\end{equation}
which can be straightforwardly evaluated after a somewhat lengthy calculation. For the sake of conciseness, the common factor is taken out
\begin{equation}\label{eq:EMfields}
\bm{E} = \sqrt{2}\frac{\mathcal{A}_0}{\sigma_\eta} e^{-r^2/\sigma_\eta^2}\bm{\mathcal{E}}, \quad \bm{B} = \sqrt{2}\frac{\mathcal{A}_0}{\sigma_\eta} e^{-r^2/\sigma_\eta^2}\bm{\mathcal{B}},
\end{equation}
where the remaining dimensionless components of the electric field are
\begin{align}
	\label{eq:Ex}\mathcal{E}_x &= \frac{r}{\omega_0\sigma_\eta^2}\left[\frac{2xy}{\sigma_\eta\sigma_0}\sin \Psi_1(0,2) - \cos \Psi_1(2,1) \right],\\
	\label{eq:Ey}\mathcal{E}_y &= \frac{r\sigma_0}{\sigma_\eta}T_1(2) + \frac{1}{\omega_0\sigma_\eta}T_2,\\
	\label{eq:Ez}\mathcal{E}_z &= \frac{\sigma_0}{\sigma_\eta}\left[\frac{2ry}{\sigma_\eta\sigma_0}\cos \Psi_1(0,1) + \sin \Psi_1(1,0) \right].
\end{align}
The phases are defined in Eq. (\ref{eq:phase}). Similarly, the dimensionless components of the magnetic field are given by
\begin{align}
	\label{eq:Bx}\mathcal{B}_x &= \frac{r\sigma_0}{\sigma_\eta}T_1(1)  + \frac{1}{\omega_0\sigma_\eta}T_2,\\
	\label{eq:By}\mathcal{B}_y &= -\frac{r}{\omega_0\sigma_\eta^2}\left[\frac{2xy}{\sigma_\eta\sigma_0}\sin \Psi_1(0,2) - \cos \Psi_1(2,1) \right],\\
	\label{eq:Bz}\mathcal{B}_z &= -\frac{\sigma_0}{\sigma_\eta}\left[\frac{2rx}{\sigma_\eta\sigma_0}\cos \Psi_1(0,1) - \cos \Psi_1(1,0) \right],	
\end{align}
where
\begin{equation}
\begin{split}
	T_1(j) &= \left[(-1)^j\omega_0 - \frac{r^2}{\eta_0\sigma_\eta^2}\right] \sin \Psi_1(0,0) \\
	&+ \frac{\sigma_0}{\eta_0 \sigma_\eta}\left[\sin \Psi_1(0,1) - \frac{2r^2\eta}{\sigma_\eta^2\eta_0}\cos\Psi(0,1)\right],
\end{split}
\end{equation}
\begin{equation}
	T_2 = \frac{2ry^2}{\sigma_\eta^2\sigma_0}\sin \Psi_1(0,2) - \frac{2y}{\sigma_\eta}\cos \Psi_1(1,1)\,.
\end{equation}
 
In order to derive these expressions the trigonometric angle addition formulas and identities
\begin{align}
\sin\left[\arctan\left(\frac{\eta}{\eta_0}\right)\right] &= \frac{\sigma_0 \eta}{\sigma_\eta \eta_0}\,,\\
\cos\left[\arctan \left(\frac{\eta}{\eta_0}\right)\right] &= \frac{\sigma_0}{\sigma_\eta}
\end{align}
were frequently used. 
\section{Average $\ell = 1$ OAM FF beam power}\label{sec:beam_power}
In this appendix, a formula for the average beam power in the $\ell = 1$ OAM FF beam is obtained. The cycle average for a general function $f(\eta,r,\theta,\phi)$ can be written as
\begin{equation}\label{eq:time_average}
	\overline{f(\eta,r,\theta)} = \frac{1}{2\pi}\int^{2\pi}_0 f(\eta,r,\theta, \phi) d\phi\,,
\end{equation}
where the average is calculated over one cycle. In this work, the averages are performed for an ultra-relativistic observer who is approximately co-moving with the field focus $\eta = t + z \approx 0$. At focus, the phase defined in Eq. (\ref{eq:phase}) can be written as
\begin{equation}
	\Psi_1(a)|_{\eta = 0} = \omega_0 \phi  + (1-a) \theta\,.
\end{equation}
The cycle averages of the following expressions [see definition Eq. (\ref{eq:time_average})] are useful
\begin{align}
	\overline{\sin [\Psi_1(a_1)]\sin[\Psi_1(a_2)]}|_{\eta = 0} &= \frac{1}{2}\cos[(a_1-a_2)\theta],\\
	\overline{\cos [\Psi_1(a_1)]\cos[\Psi_1(a_2)]}|_{\eta = 0} &= \frac{1}{2}\cos[(a_1-a_2)\theta],\\
	\overline{\sin [\Psi_1(a_1)]\cos[\Psi_1(a_2)]}|_{\eta = 0} &= \frac{1}{2}\sin[(a_1-a_2)\theta]\,.
\end{align}

The average power transmitted through the $xy$ plane at $\eta = 0$ is given by the cycle-averaged Poynting vector flux
\begin{equation}
	P_\text{ave} = \int dx dy \left.\overline{E_xB_y - E_yB_x}\right|_{\eta = 0}.
\end{equation}
In the simulations presented in Section \ref{sec:simulations}, $\omega_0 \sigma_0 = 6\pi$, thus only the leading order terms  $\propto (\omega_0\sigma_0)^n$ are considered. The leading-order term has $n = 2$, there is no contribution with $n=1$, and any terms with $n \leq 0$ are neglected. Ultimately, the leading contribution to the beam power comes from
\begin{equation}
	\left. \overline{T_1(2)T_1(1)}\right|_{\eta = 0} \approx - \frac{1}{2}\omega_0^2\,.
\end{equation}
After performing the integration over the transverse coordinates the final expression for the average power reads
\begin{equation}\label{eq:power}
	\begin{split}
		P_\text{ave} = \frac{\pi}{4}\mathcal{A}_0^2 &\left[\omega_0^2 \sigma_0^2 + O(1)\right] \\
		&\approx 21.5[\text{GW}] \left(\xi_0 \frac{\sigma_0}{\lambda_0} \right)^2
	\end{split}
\end{equation}
and is identical to the conventional LG10 beam. The numerical value in the last expression is given for the field strength $\xi_0$ scaled to electron (positron) mass and charge. In the cases considered here, $\sigma_0 = 3 \lambda_0$, which means that the pulses with $t_\text{int} = 20$ ps and $\xi_0 = 5$ have a power of about 5 TW and a total energy of about 200 J, see Eq. (\ref{eq:totenergy}). 

\section{Transverse ponderomotive force}\label{sec:ponderomotive_app}
A formula for the transverse ponderomotive force acting on a charged particle in the FF pulse is derived in this appendix. Radiation-reaction is neglected for this derivation. In terms of the vector four-potential, the Lorentz equation is
\begin{equation}
	\frac{d}{d\tau}(m u^\mu) = q (\partial^\mu A^\nu - \partial^\nu A^\mu) u_\nu\,.
\end{equation}
Since $u_\nu \partial^\nu = d/d\tau$ along the particle trajectory, the second term on the RHS can be combined with the proper time derivative on the LHS. The remaining product $u_\nu A^\nu$ on the RHS can be expressed in lightcone coordinates, yielding
\begin{equation}\label{eq:LFlightcone}
	\begin{split}
		\frac{d}{d\tau}&(mu^\mu + qA^\mu) \\
		&= \frac{q}{2}\left(u_-\partial^\mu A_+ + u_+ \partial^\mu A_-\right) - q\bm{u}_\perp \cdot \partial^\mu \bm{A}_\perp\,.
	\end{split}
\end{equation}
In this expression the definition of the dot product of two four-vectors $a^\mu$ and $b^\mu$ in the lightcone coordinates
\begin{equation}
	a_\mu b^\mu = \frac{1}{2}(a_+ b_- + a_- b_+) - \bm{a}_\perp \cdot \bm{b}_\perp\,,
\end{equation}
was used. Finally, the lightcone components $a_+$ and $a_-$ are defined as
\begin{equation}\label{eq:lightcone}
	a_+ = a^0 + a^z, \quad a_- = a^0 - a^z\,.
\end{equation}

In the gauge used here $A_+$ vanishes, and therefore the first term on the RHS of Eq. (\ref{eq:LFlightcone}) does not contribute. From the constraint $u^2 = 1$ on the four-velocity, $u_+$ can be expressed as 
\begin{equation}
	u_+ = \frac{1+\bm{u}_\perp^2}{u_-} \approx \frac{1}{2\gamma}
\end{equation}
provided that the perpendicular velocity is small ($\xi_0 \ll \gamma$), and the particle moves with ultra-relativistic velocity in the negative $z$ direction ($u_- \approx 2\gamma$). This allows the second term on the RHS of Eq. (\ref{eq:LFlightcone}) to be neglected compared to the third term. Employing this approximation, one finds
\begin{equation}\label{eq:LFsimple}
	\frac{d}{d\tau}(mu^\mu + qA^\mu) \approx  - q\bm{u}_\perp \cdot \partial^\mu \bm{A}_\perp\,.
\end{equation}
In the perpendicular direction, the ansatz
\begin{equation}\label{eq:ansatz}
	\bm{u}_\perp = - \frac{q}{m} \bm{A}_\perp + \delta \bm{u}_\perp\
\end{equation}
can be made, where the first term is the exact solution of perpendicular motion in a plane wave \cite{Landau_b_2_1975}, and the second term represents a deviation from this motion due to nontrivial transverse structure of the field. Upon substituting this ansatz in to both sides of Eq. (\ref{eq:LFsimple}), the leading-order correction to the perpendicular component of the four-velocity reads
\begin{equation}\label{eq:result}
	\frac{d}{d\tau}\delta \bm{u}_\perp \approx - \frac{q^2}{2m^2} \bm{\nabla}_\perp A_\perp^2\,,  
\end{equation}
where the identity $\bm{A}_\perp \cdot \nabla_i \bm{A}_\perp = \nabla_i A_\perp^2 / 2$ was used. In a plane wave, $\bm{A}_\perp$ does not depend on the transverse coordinates and this correction vanishes as expected. Finally, the proper-time derivative can be written in terms of the derivative with respect to the laboratory time because $d/d\tau = \gamma d/dt$ along the particle trajectory. 

Upon applying the cycle-averaging procedure [defined in Eq. (\ref{eq:time_average})] at focus ($\eta = 0$) to Eq. (\ref{eq:ansatz}), the oscillatory plane wave term vanishes [see its prescription in Eq. (\ref{eq:aperp})] and one obtains
\begin{equation}
	\overline{\bm{u}_\perp} = \overline{\delta \bm{u}_\perp}\,.
\end{equation}
In order to carry out the cycle-averaging integration, the functions $r(t)$ and $\theta(t)$, corresponding to the polar coordinates of the charge at the time $t$ on the $xy$ plane, are considered to be changing on a much slower scale and effectively constant in the averaging over one cycle. This result implies that the cycle-averaged perpendicular velocity is solely given by the term describing the deviation from the plane wave motion. Therefore performing a cycle average of Eq. (\ref{eq:result}) yields
\begin{equation}\label{eq:ponderomotive_app}
	\frac{d\overline{\bm{u}}_\perp}{dt} \approx - \frac{q^2}{2m^2\gamma_0} \bm{\nabla}_\perp \overline{A_\perp^2}|_{\eta = 0}\,,
\end{equation}
where it was assumed that the relativistic gamma factor of the particle remains approximately unchanged and can be taken out of the average, which is correct up to terms on the order of $O(1/\gamma_0)$ \cite{DiPiazza:2013vra}.
\section{Normalized transverse emittance}\label{sec:emittance_def}
The transverse emittance is defined as a quantity proportional to the phase-space area of a bunch in the transverse direction. For computational purposes, the statistical definition of normalized transverse emittance is more useful \cite{floettmann2003some}. By generalizing the standard definition of the emittance to two-dimensional vectors, one obtains
\begin{equation}\label{eq:emittance}
	\varepsilon_\perp = \frac{1}{m}\sqrt{\sigma_{\bm{r}}^2 \sigma_{\bm{p}_\perp}^2 - \sigma_{\bm{r, p}_\perp}^4 }\,.
\end{equation}
The variance $\sigma_{\bm{r}}^2$ of the transverse position vector $\bm{r} = (x,y)$ is defined as
\begin{equation}
	\sigma_{\bm{r}}^2 = \langle \bm{r} \cdot \bm{r} \rangle - \langle \bm{r}\rangle \cdot \langle \bm{r} \rangle \,,
\end{equation}
where, in polar coordinates, 
\begin{equation}\label{eq:r2var}
	\langle \bm{r} \cdot \bm{r} \rangle = \langle r^2 \rangle\,.
\end{equation}
The relativistic transverse momentum is $\bm{p}_\perp = m \gamma (\beta_x, \beta_y)$. Its variance $\sigma_{\bm{p}_\perp}^2$ is 
\begin{equation}
	\sigma_{\bm{p}_\perp}^2 = \langle \bm{p}_\perp \cdot \bm{p}_\perp \rangle - \langle \bm{p}_\perp \rangle \cdot \langle \bm{p}_\perp \rangle \,,
\end{equation}
where
\begin{equation}\label{eq:pr}
	\bm{p}_\perp \cdot \bm{p}_\perp = m^2\gamma^2 (r'^2 + r^2 \theta'^2)\,.
\end{equation}
Finally, the cross variance $\sigma_{\bm{r, p}_\perp}$ is given by
\begin{equation}
	\sigma_{\bm{r, p}_\perp}^2 = \langle \bm{r}\cdot \bm{p}_\perp \rangle - \langle \bm{r} \rangle \cdot \langle \bm{p}_\perp \rangle\,,
\end{equation}  
where 
\begin{equation}\label{eq:rpperp}
	\langle \bm{r} \cdot \bm{p}_\perp \rangle = m \langle \gamma x \beta_x  + \gamma y \beta_y \rangle = m \langle \gamma rr'\rangle \,.
\end{equation} 
Now, since the average transverse position $\langle \bm{r} \rangle$ and average transverse momentum $\langle \bm{p}_\perp \rangle$ are approximately zero throughout the evolution of the bunch due to cylindrical symmetry, the normalized emittance can be re-written as
\begin{equation}
	\varepsilon_\perp = \frac{1}{m}\sqrt{\langle \bm{r} \cdot \bm{r}\rangle \langle \bm{p}_\perp \cdot \bm{p}_\perp \rangle - \langle\bm{r} \cdot \bm{p}_\perp \rangle^2 }\,.
\end{equation}
After substitution from Eqs. (\ref{eq:r2var}), (\ref{eq:pr}), and (\ref{eq:rpperp}) one finds
\begin{equation}
	\varepsilon_\perp = \sqrt{\langle r^2 \rangle \langle \gamma^2r'^2 + \gamma^2 r^2 \theta'^2 \rangle - \langle \gamma r r'\rangle^2 }\,.
\end{equation}
Finally, the approximations that the motion is ultra-relativistic, that radiation reaction is neglected, and that the fields are relatively weak , i.e., $\xi_0 \ll \gamma_0$, can be made. With these approximations, the relativistic Lorentz factor $\gamma$ is approximately constant, has very little spread, and can be taken out of the ensemble averages. This gives the normalized transverse emittance
\begin{equation}\label{eq:emittance_approx}
	\varepsilon_\perp \approx \gamma \sqrt{\langle r^2 \rangle \langle r'^2 + r^2 \theta'^2 \rangle - \langle r r' \rangle^2 }\,,
\end{equation}
which appears in Eq. (\ref{eq:Rpp}) for the evolution of the RMS radius. 
\section{Coulomb repulsion among particles}\label{sec:coulomb}
In this appendix, an estimate of the Coulomb repulsion force is presented, and the conditions for which this force can be neglected compared to the ponderomotive force are established. The charged particle bunches are modeled analytically by the charge distribution in Eqs. (\ref{eq:positiondistrib_text}) and (\ref{eq:lambda_text}). As was shown in Ref. \cite{Tamburini:2019tzo}, the electric field generated in the rest frame of this distribution can be computed exactly. 

At $z =0$ (in the middle of the bunch), only a purely radial field remains
\begin{equation}
	\bm{E}(r,\theta,0) = E_r \hat{\bm{r}} = \frac{1}{4\pi}\frac{qN}{Lr}f(r)\hat{\bm{r}}\,,
\end{equation}
where $f(r)$ is given by
\begin{equation}\label{eq:fr}
	\begin{split}
		f(r) &= \frac{L}{\sqrt{L^2/4 + r^2}}\text{erf}\left(\frac{\sqrt{L^2/4+r^2}}{\sqrt{2}\sigma_r} \right)\\
		&- 2e^{-r^2/2\sigma^2_r}\text{erf}\left(\frac{L}{2\sqrt{2}\sigma_r} \right)\,.
	\end{split}
\end{equation}
In the laboratory frame, this transverse electric field is enhanced by a factor of $\gamma_0$ because of the Lorentz transformation. A magnetic field in the azimuthal direction is also induced due to presence of a non-zero current in this frame. Performing the Lorentz transformation explicitly \cite{Jackson_b_1975}, the Cartesian components of the fields are
\begin{align}
	\bm{E}_\text{lab}(x,y,0) &= \gamma_0 E_r \left(\frac{x}{r} \hat{\bm{x}} + \frac{y}{r} \hat{\bm{y}}\right)\,,\\
	\bm{B}_\text{lab}(x,y,0) &= \gamma_0 \beta_0 E_r \left( - \frac{y}{r} \hat{\bm{x}} +  \frac{x}{r} \hat{\bm{y}}\right)\,.
\end{align}
Using these fields, the laboratory frame four-force is given by
\begin{equation}
	\begin{split}
		\mathcal{F}^\mu &= q F^{\mu\nu}u_\nu = q \gamma_0 \gamma E_r\\
		&  \times \left[\beta_r, \frac{x}{r}(1-\beta_0 \beta_z), \frac{y}{r}(1-\beta_0\beta_z), \beta_0 \beta_r\right]\,.
	\end{split}
\end{equation}
Assuming that the particle deviates only slightly from the ultra-relativistic straight-line motion, i.e., $\beta_0 \approx \beta_z$ and $\gamma \approx \gamma_0 = (1-\beta_0^2)^{-1/2}$, the components of the force in the laboratory frame are
\begin{equation}
	\begin{split}
		\mathcal{F}^0 = q \gamma_0^2 E_r \beta_r, &\quad \mathcal{F}^r = q E_r,\\ \mathcal{F}^\theta = 0, &\quad \mathcal{F}^z = q \gamma_0^2 E_r \beta_0 \beta_r\,.
	\end{split}
\end{equation}
Under the same approximations (neglecting time derivatives of $\gamma = \gamma_0$), the equation for the radial motion in the absence of the fields of the FF pulse is
\begin{equation}\label{eq:accel_coulomb}
	r''  - \frac{L_z^2}{m^2 \gamma_0^2 r^3} = \frac{qE_r}{m\gamma^2_0} = \frac{1}{4\pi}\frac{q^2 N}{mLr\gamma_0^2}f(r)\,.
\end{equation}
Notice that the dependence on the gamma factor $\gamma_0$ is the same as for the ponderomotive force Eq. (\ref{eq:radial}). 

Now, the acceleration arising from the ponderomotive force [Eq. (\ref{eq:radial})] and the acceleration due to Coulomb repulsion (\ref{eq:accel_coulomb}) can be compared. Since both of these forces are zero on axis and increase with radius up to a certain point, it makes sense to compare the accelerations at their respective maxima. For the ponderomotive force, this is at the radius
\begin{equation}\label{eq:RP}
	r_{P,\text{max}} = \frac{\sqrt{5-\sqrt{17}}}{2}r_\text{max}\,,
\end{equation}
which can be derived by taking the second derivative of the ponderomotive potential in Eq. (\ref{eq:potential}) and solving for the root in the interval $(0,r_\text{max})$. The acceleration due to ponderomotive force at its maximum reads
\begin{equation}
	a_P(r=r_{P,\text{max}}) = 0.21 \frac{\xi_0^2}{\gamma_0^2 \sigma_0}\,.
\end{equation}
The radius of the maximum Coulomb force $r_{C,\text{max}}$ must be determined numerically. The electron bunch is typically much longer than it is wide $L > \sigma_r$, which allows the error functions in Eq. (\ref{eq:fr}) to be approximated by 1. The remaining function $\sigma_r f(r)/r$ has an upper bound $\sigma_r f(r) / r < 1$ for any $L > \sigma_r$. Substituting this in to the expression for the acceleration in Eq. (\ref{eq:accel_coulomb}), one finds that the Coulomb acceleration is less than 
\begin{equation}
	a_C(r=r_{C,\text{max}}) < \frac{1}{4\pi} \frac{q^2 N}{m\gamma_0^2 L\sigma_r}\,.
\end{equation} 
Thus, the ratio of the maximal accelerations is proportional to
\begin{equation}
	\begin{split}
	\frac{a_C(r=r_{C,\text{max}})}{a_P(r=r_{P,\text{max}})} = &\frac{q^2}{4\pi m} \frac{\sigma_0}{L \sigma_r} \frac{N}{0.21\xi_0^2}\\
	&= 1.3\times 10^{-8} [\mu\text{m}] \frac{\sigma_0}{L\sigma_r} \frac{N}{\xi_0^2}\,,
	\end{split}
\end{equation}
where the numerical factor is given for electrons. For the electron bunches considered here $\sigma_0 / [L(0)\sigma_r(0)] = 0.4\ \mu\text{m}^{-1}$ and the accelerations become comparable, for example, when $\xi_0 = 10$ and $N = 2\times 10^{10}$, corresponding to a total bunch charge of 3 nC.

\hspace{1cm}

%
\end{document}